\documentclass[11pt,a4paper]{article}

\usepackage{jheppub}
\hypersetup{hidelinks = true, }
\usepackage{amsmath,amssymb}
\usepackage{enumerate}
\usepackage{amsfonts}

\usepackage{subfigure}

\usepackage{allpurpose}

\usepackage{tikz}

\usepackage[marginal,perpage,symbol]{footmisc}

\usepackage{slashed} %

\graphicspath{{./figures/}}
\usepackage{caption}
\everymath{\displaystyle}
\makeatletter
\gdef\@fpheader{}
\makeatother

\usepackage[nottoc,notlof,notlot]{tocbibind}

\usepackage{indentfirst}

\begin{document}

\title{Spin precession in the strong deflection limit}
\date{\today}

\author{Jiafei Geng,}

\author{Yunchuan Xiang,}
\emailAdd{xiang{\_}yunchuan@yeah.net}

\author[*]{Qingquan Jiang\note[*]{The corresponding author.}}
\emailAdd{qqjiangphys@yeah.net}

\author[\dag]{and Xiankai Pang\note[\dag]{The corresponding author.}}
\emailAdd{xkpang@cwnu.edu.cn}

\affiliation{\normalfont \normalfont School of Physics and Astronomy, China West Normal University, Nanchong, 637009, China}

\abstract{
The strong deflection limit (SDL) of the deflection angle is well established for general spherically symmetric spacetimes, but a systematic SDL treatment of spin precession has been lacking. We derive the SDL expansion for the spin precession angle of particles propagating along geodesics in static, spherically symmetric, and asymptotically flat spacetimes. The SDL coefficients are obtained, and a simple relation linking the precession angle to the deflection one is established. Applying the formalism to Schwarzschild and Reissner-Nordstr\"om (RN) spacetimes, we obtain fully analytic SDL coefficients for the former and perturbative charge corrections to $\mathcal{O}(Q^2)$ for the latter. These results explicitly verify the universal spin-flip of massless particles in backward scattering, which is the underlying mechanism governing the absence of the glory spot. While the spin precession of massless particles is charge-independent, massive particles acquire a non-trivial charge dependence that distinguishes the RN case from Schwarzschild.
}


\maketitle

\section{Introduction}
\label{sec:introduction}

In the standard geometrical optics (or WKB) approximation, wave propagation in curved spacetime reduces to the motion of massless particles along null geodesics~\cite{Landau:1982dva,Straumann:1984grr}. At leading order, this approximation yields the same geodesic equation regardless of the spin of the underlying field~\cite{Stone:2014fja,Oancea:2022utx}, making the scattering of a massless scalar field ($s=0$) indistinguishable from that of an electromagnetic or gravitational wave ($s=1$ or $s=2$). This spin-blindness, however, fails dramatically in the backward direction, as partial wave analysis reveals that while scalar fields exhibit a glory spot~\cite{Ford:1959sd,Zhang:1984vt,DeWittMorette:1984pk,Matzner:1985rjna}, the backward cross section for massless fields with non-vanishing spin vanishes~\cite{Zhang:1984vt,Futterman:1988sbh,Dolan:2006vja,Crispino:2009xt}. This striking difference has been investigated extensively for scalar~\cite{Crispino:2009kia,Macedo:2015qma,Huang:2020bdf}, electromagnetic~\cite{Crispino:2009xta}, and fermionic~\cite{Dolan:2006vja} fields scattered by various black hole spacetimes.

Classically, the absence of the glory spot reflects a perfect cancellation arising from spin precession along orbits in mutually orthogonal planes~\cite{Crispino:2009xta}. 
Concretely, for massless particles, a spin vector confined to the equatorial plane reverses its orientation upon backward scattering along equatorial geodesics, whereas it remains invariant along geodesics lying in the orthogonal plane~\cite{Pang:2024tco}. 
Quantitatively, this behavior is encoded in the parallel transport of the spin vector, parameterized by the precession angle~$\chi$~\cite{Kubiznak:2008zs,Connell:2008vn}.
At subleading order in the WKB approximation, the spin parallel-transport equation reduces to a single differential equation for~$\chi$, whose integral form closely mirrors that of the deflection angle~\cite{Pang:2024tco}.

Backward scattering necessarily involves large deflection angles, the particle must wind around the central object before returning toward the source, and therefore probes the strong deflection regime. This regime has acquired renewed observational significance in recent years, driven by the LIGO--Virgo direct detection of gravitational waves from binary black hole mergers~\cite{LIGOScientific:2016aoc,LIGOScientific:2018mvr} and the Event Horizon Telescope's imaging of supermassive black hole shadows in M87$^*$~\cite{Akiyama:2019bqs,EventHorizonTelescope:2019dse} and Sgr~A$^*$~\cite{EventHorizonTelescope:2022exc,EventHorizonTelescope:2022xqj}, which together have ushered in an era of direct strong-field gravity tests.

It is well established that particle deflection diverges logarithmically in the \acrfull{sdl} when expanded near the critical radius $r_c$ in spherically symmetric spacetimes~\cite{Darwin:1959gfp,Virbhadra:1999nm,Claudel:2000yi,Bozza:2002zj,Iyer:2006cn,Tsupko:2014wza,Pang:2018jpm,Feleppa:2024kio}. Nevertheless, the existing literature has largely neglected the spin precession in this regime. This manuscript addresses this gap by deriving the \acrshort{sdl} expansion for the precession angle, proving its logarithmic divergence similar to that of the deflection angle. By applying these results to Schwarzschild and \acrfull{rn} spacetimes, we extract explicit \acrshort{sdl} coefficients and analyze the transition of spin precession from non-relativistic to extreme relativistic domains.

Our work contributes to the broader effort of characterizing spinning particle dynamics in curved spacetime.
The spin degrees of freedom are encoded in a 4-vector~$S^\mu$, orthogonal to the 4-velocity~$u^\mu$~\cite{Weinberg:1972kfs,Khriplovich:2008ni}.
At leading order in the WKB approximation, $S^\mu$ is simply parallel-transported along the geodesic~\cite{Dolan:2017zgu}.
Beyond this adiabatic limit, spin--orbit coupling induces deviations from geodesic motion, governed by the \acrfull{mpd} equations~\cite{Mathisson:2010opl,Papapetrou:1951pa,Rudiger:1981uu}.
This formalism has been extensively applied to circular orbits~\cite{Mohseni:2010rm,Zhang:2017nhl}, spin--spin interactions~\cite{Iorio:2011ubn,Chakraborty:2016mhx}, and analytic solutions in spherically symmetric spacetimes~\cite{Witzany:2023bmq}. 
Very recently, it was shown that the \acrshort{mpd} equations at linear order are separable~\cite{Witzany:2026eqc} in spacetimes with hidden symmetries~\cite{Frolov:2017kze}.
For massless particles, the \acrshort{mpd} framework requires careful treatment of spin supplementary conditions~\cite{Poisson:2011nh,Duval:2016hxo}.
Complementarily, the spinoptics formalism has been developed to systematically incorporate helicity-dependent corrections beyond geometric optics, with applications to both electromagnetic~\cite{Frolov:2020uhn,Shoom:2020zhr,Frolov:2024ebe,Takeuchi:2026pyi} and gravitational waves~\cite{Andersson:2020gsj,Frolov:2024qow}.
These developments have enabled detailed analyses of the gravitational spin Hall effect and spin Faraday rotation~\cite{Harte:2018wni,Oancea:2020khc,Oancea:2023ylb}.
In the present analysis, however, we restrict our attention to the WKB regime where the spin vector is parallel-transported along null geodesics. 
This controlled simplification retains the essential physics of backward scattering while rendering the problem sufficiently tractable to yield fully analytic \acrshort{sdl} coefficients.

The paper is organized as follows. In Sec.~\ref{sec:deflectionangle}, we briefly review the \acrshort{sdl} for the deflection angle in a general spherically symmetric spacetime,  to set notation and conventions. Sec.~\ref{sec:spinpreceession} develops the core theoretical results, deriving the \acrshort{sdl} expansion for the spin precession angle, identifying its logarithmic divergence, and establishing a direct link to the deflection angle. We then apply this formalism in Sec.~\ref{sec:examples} to Schwarzschild and \acrshort{rn} backgrounds, providing explicit analytic coefficients and comparing them with exact results. Sec.~\ref{sec:discussion} summarizes our findings and outlines directions for future work.

\section{The strong deflection limit of the deflection angle}
\label{sec:deflectionangle}
We begin with a brief review of the deflection angle in the \acrshort{sdl}~\cite{Bozza:2002zj,Feleppa:2024kio}. %
Considering the following metric of a general spherically symmetric spacetime,
\begin{IEEEeqnarray}{rCl}
	\dd s^2 &=& f(r)\dd t^2-\frac{1}{f(r)}\dd r^2 -r^2\dd\theta^2-r^2\sin^2\theta\dd\varphi^2 \quad, \label{eq:sphericalmetric}
\end{IEEEeqnarray} 
we further require it's asymptotically flat, i.e., $f(r)\to1$ as $r\to\infty$. %
The orbit equation governing the particle deflection can be obtained by solving the \acrfull{hjeq} equation~\cite{Landau:mechanics,Landau:1982dva}. In a static and spherically symmetric spacetime given by the metric \eqref{eq:sphericalmetric}, the particle will move in a plane which can be chosen as the one with $\theta=\pi/2$. And there are two conserved quantities, the energy $E$ and the angular momentum $L$ per unit mass of the particle, which can be written in the form of velocity $v$ and impact parameter $b$ at infinity as follows  
\begin{IEEEeqnarray}{rCl}
	E &=& \frac{1}{\sqrt{1-v^2}} \quad, \quad L=\frac{bv}{\sqrt{1-v^2}}=\frac{\eta}{\sqrt{1-v^2}}\quad , \label{eq:ELbveta}
\end{IEEEeqnarray} 
where we have introduced a new parameter $\eta=bv$ for convenience, in particular for slow particles since we are considering both massive and massless ones. In the non-relativistic limit with $v\to0$, $\eta$ remains finite while the impact parameter $b$ itself diverges~\cite{Pang:2018jpm}. And for massless particles, we have $L/E=\eta=bv=b$.

\paragraph{The orbit equation.} Since the spacetime is static and spherically symmetric,  $t$ and $\varphi$ are cyclic variables and can be separated in the action of a particle moving in the equatorial plane~\cite{Landau:1982dva}
\begin{IEEEeqnarray}{rCl}
	S &=& -E t +L \varphi +S_r(r,L,E) \quad, \label{eq:actionELSr}
\end{IEEEeqnarray} 
where $S_r(r,L,E)$ is an unknown function independent of $t$ and $\varphi$ and to be determined by the \acrshort{hjeq} equation~\cite{Landau:1982dva,Pang:2018jpm}:
\begin{IEEEeqnarray}{rCl}
	\frac{E^2}{f(r)}-\frac{L^2}{r^2}-f(r)\left[\frac{\dd S_r(r)}{\dd r}\right]^2 &=& \kappa \quad, \label{eq:HJeqSr}
\end{IEEEeqnarray} 
where $\kappa=0,~1$ for massless and massive particles respectively. Again, since $\varphi$ is a cyclic variable, we have~\cite{Landau:mechanics}
\begin{IEEEeqnarray*}{rCl}
	\frac{\partial S}{\partial L} &=& \text{constant} \quad, 
\end{IEEEeqnarray*} 
and therefore
\begin{IEEEeqnarray}{rCl}
	\frac{\partial}{\partial r}\left(\frac{\partial S}{\partial L}\right) &=& 0 \quad. \label{eq:prpLS}
\end{IEEEeqnarray} 
Solving $S_r(r,L,E)$ from the \acrshort{hjeq} equation \eqref{eq:HJeqSr} and substituting into Eq.~\eqref{eq:prpLS}, we can obtain the orbit equation. In terms of $v$ and $\eta$, the deflection angle satisfies 
\begin{IEEEeqnarray}{rCl}
	\frac{\dd\varphi}{\dd r} &=& \frac{L}{r^2}\frac{1}{\sqrt{E^2-V_{\text{eff}}(r)}}=\frac{\eta}{r^2}\frac{1}{\sqrt{1-(1-v^2)V_{\text{eff}}(r)}} \quad, \label{eq:orbitequationdphidr}
\end{IEEEeqnarray} 
where the effective potential $V_{\text{eff}}$ is defined as
\begin{IEEEeqnarray}{rCl}
	\left(\frac{\mathrm{d} r}{\mathrm{d} \tau}\right)^2+V_{\text{eff}}=E^2  \quad, 
\end{IEEEeqnarray} 
where $\tau$ is the affine parameter along the geodesic.
More explicitly, the effective potential can be written as %
\begin{IEEEeqnarray}{rCl}
	V_{\text{eff}} &=& \left(\frac{L^2}{r^2}+\kappa\right)f(r)=\left[\frac{\eta^2}{r^2(1-v^2)}+\kappa\right]f(r) \quad. \label{eq:Veffkappa}
\end{IEEEeqnarray} 

For massive particles, we have $\kappa=1$ and the orbit equation becomes 
\begin{IEEEeqnarray}{rCl}
	\frac{\dd\varphi}{\dd r} &=&\frac{\eta}{r}\frac{1}{\sqrt{r^2-\left[\eta^2+(1-v^2)r^2\right]f(r)}}  \quad. \label{eq:orbitequationdphidrmassive}
\end{IEEEeqnarray} 
The massless case can be obtained either by setting $\kappa=0$ in the effective potential~\eqref{eq:Veffkappa} and then substituting it into the orbit equation \eqref{eq:orbitequationdphidr}, or taking $v\to 1$ limit in the massive orbit Eq.~\eqref{eq:orbitequationdphidrmassive}~\cite{Pang:2018jpm}.%
For simplicity, we will focus on the massive particles (whose motion is governed by the massive orbit equation \eqref{eq:orbitequationdphidrmassive}) in the following discussion, and take the $v\to1$ limit for massless particles when necessary. %

\paragraph{The deflection angle.}
For the deflection of a particle around the black hole, let's denote $r_0$ the radius of the closest approach point in the orbit. For each speed $v$, there is a critical closest approach $r_c$, such that the deflection angle diverges~\cite{Tsupko:2014wza}, 
and correspondingly the parameter $\eta$ has the critical value $\eta_c$ as well.
To determine $r_c$, we need two conditions $\frac{\dd r}{\dd \varphi}=0$ and $\frac{\dd^2 r}{\dd\varphi^2}=0$, which translate to the following two equations 
\begin{IEEEeqnarray}{rCl}
	E^2-V_{\text{eff}}(r_c) &=& \frac{1}{1-v^2}-V_{\text{eff}}(r_c)=0 \quad,\quad \left.\frac{\dd}{\dd r}V_{\text{eff}}\right|_{r=r_c}=0\quad. \label{eq:rcdefeq}
\end{IEEEeqnarray} 

\begin{figure}[htpb]
	\centering
	\includegraphics[width=0.8\textwidth]{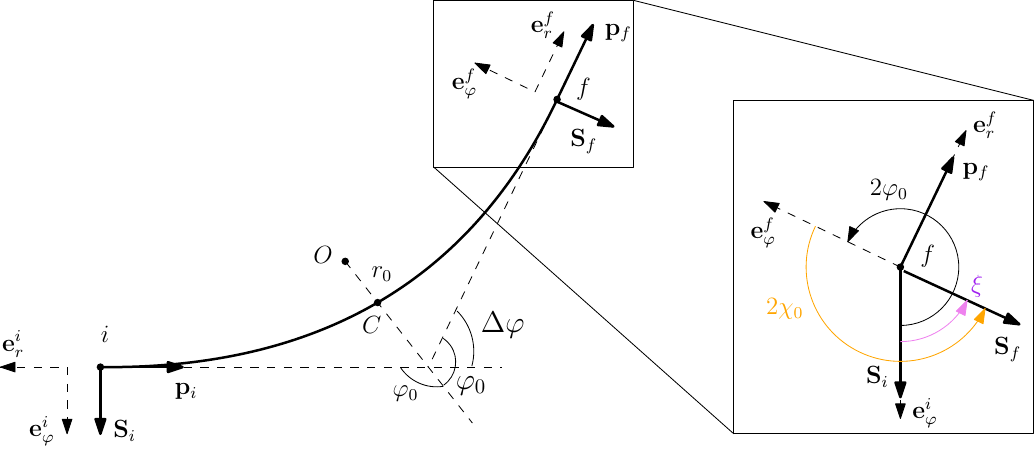}
	\caption{Deflection and spin precession of a particle scattered by a central black hole.
    $O$ denotes the black hole center; $i$ and $f$ label the initial and final points of the scattered particle (both located at spatial infinity in practice).
    $r_0$ is the distance of closest approach from $O$ to the particle orbit $C$.
    $\vec{p}_i$ and $\vec{S}_i$ are the momentum and spin of the particle at $i$, with $\vec{S}_i$ defined in the particle's rest frame.
    $\vec{e}_r^{\,i}$ and $\vec{e}_\varphi^{\,i}$ are unit vectors in the directions of increasing $r$ and $\varphi$ at $i$; $\vec{p}_f$, $\vec{S}_f$, $\vec{e}_r^{\,f}$, $\vec{e}_\varphi^{\,f}$ are the corresponding quantities at $f$.
    The angles between $\vec{e}_\varphi^{\,f}$ and $\vec{e}_\varphi^{\,i}$, between $\vec{S}_f$ and $\vec{e}_\varphi^{\,f}$, and between $\vec{S}_f$ and $\vec{S}_i$ are $2\varphi_0$, $2\chi_0$, and $\xi$, respectively.}
	\label{fig:anglephichixi}
\end{figure}

Let $\varphi_0=\varphi(r_0)$ to be the azimuthal angle when the particle approaches to $r_0$, then the deflection angle can be written as 
\begin{IEEEeqnarray}{rCl}
	\Delta\varphi &=& 2\varphi_0-\pi \quad,  \label{eq:DeltaPhideflectionangle}
\end{IEEEeqnarray} 
as shown in Fig.~\ref{fig:anglephichixi}.

\paragraph{The strong deflection limit.}

It's well known that in the \acrshort{sdl}, i.e., when $r_0\to r_c$\footnote{Equivalently, we have $\eta\to\eta_c$. As shown in \ref{sec:expr0andeta}, we have that if $r_0=r_c(1+\delta)$ with $\delta\to0$, then $\eta=\eta_c(1+\gamma \delta^2)$, where $\gamma$ is a constant given by equation \eqref{eq:r0etagamma}.}, the deflection angle $\Delta\varphi$ diverges logarithmically~\cite{Bozza:2002zj,Tsupko:2014wza,Liu:2015zou,Pang:2018jpm,Feleppa:2024kio}. 
More specifically, let $r_0=r_c(1+\delta)$ for $\delta\to0$, $\varphi_0$ has the following divergence structure~\cite{Bozza:2002zj,Feleppa:2024kio}
\begin{IEEEeqnarray}{rCl}
	\varphi_0 &=&-\lambda_{\varphi}\ln\delta+\zeta_{\varphi,D}+\zeta_{\varphi,R}  \quad, \label{eq:phi0divergencedelta}
\end{IEEEeqnarray} 
where $\lambda_{\varphi},~\zeta_{\varphi,D}$ and $\zeta_{\varphi,R}$ are constants independent of $\delta$.

To show the logarithmic divergent structure given by Eq.~\eqref{eq:phi0divergencedelta}, we introduce a new variable $z$ such that \cite{Bozza:2002zj,Feleppa:2024kio} 
\begin{IEEEeqnarray}{rCl}
	z &=& \frac{f(r)-f(r_0)}{1-f(r_0)} \quad.  \label{eq:varzdef}
\end{IEEEeqnarray} 
We see that $z\to0$  as $r\to r_0$ and $z\to1$ as $r\to\infty$.
Eq.~\eqref{eq:phi0divergencedelta} can be shown by first transforming the orbit equation \eqref{eq:orbitequationdphidrmassive} into the new variable $z$ and then extracting the logarithmic divergence behaviour. For which we need two differentials\footnote{These two differentials can be obtained by noting that $\frac{\dd r}{\dd z}=\frac{1}{\dd z/\dd r}$.}
\begin{IEEEeqnarray}{rCl}
	\frac{\mathrm{d} r}{\mathrm{d} z} &=& \frac{1-f_0}{f'(r)} \quad, \quad \frac{\mathrm{d}^2 r }{\mathrm{d} z^2}=-\frac{(1-f_0)^2f''(r)}{f'(r)^3} \quad.
\end{IEEEeqnarray} 
Then the orbit equations \eqref{eq:orbitequationdphidrmassive} transforms to
\begin{IEEEeqnarray}{rCl}
	\frac{\mathrm{d} \varphi}{\mathrm{d} z}	 &=& \frac{\mathrm{d} \varphi}{\mathrm{d} r}\frac{\mathrm{d} r}{\mathrm{d} z}=\frac{R_{\varphi}(z,r_0,v)}{\sqrt{P(z,r_0,v)}} \quad. 
\end{IEEEeqnarray} 
The functions $R_{\varphi}(z,r_0,v)$ and $P(z,r_0,v)$ are defined as follows 
\begin{IEEEeqnarray}{rCl}
	R_{\varphi}(z,r_0,v) &=& \frac{r_0(1-f_0)\sqrt{1-f_0(1-v^2)}}{r^2f'[r(z)]}  \quad, \label{eq:defRvarphi}  \\ 
	P(z,r_0,v)&=& f_0+\frac{r_0\left[1-f_0(1-v^2)\right]}{r(z)^2} \quad,  \label{eq:defP}
\end{IEEEeqnarray} 
where we have replaced $\eta$ using its relation~\eqref{eq:etar0relation} with $r_0$, since we are going to expand the deflection angle with respect to $r_0=r_c(1+\delta)$ rather than $\eta=\eta_c(1+\varepsilon)$\footnote{The relation between these two kinds of expansions is given in appendix~\ref{sec:expr0andeta}.}.
It's worth to emphasize that in Eqs.~\eqref{eq:defRvarphi} and~\eqref{eq:defP}, $r(z)$ should be understood as a function of $z$, and consequently $R_\varphi$ and $P$ are functions of $z$ instead of $r$.

The angle $\varphi_0$ can then be given by an integral
\begin{IEEEeqnarray}{rCl}
	\varphi_0 &=& \int_0^1\dd z \frac{R_\varphi(z,r_0,v)}{\sqrt{P(z,r_0,v)}} \quad. \label{eq:phi0integral}
\end{IEEEeqnarray} 

Near $z\to0$, the function $P(z,r_0,v)$ can be expanded as $P(z,r_0,v)=\alpha z +\beta z^2$, where $\alpha$ and $\beta$ are constants
\begin{IEEEeqnarray}{rCl}
	\alpha &=& \frac{(1-f_0)\left[2f_0-2f_0^2(1-v^2)-r_0f'_0\right]}{r_0f'_0} \quad,\label{eq:alphar0relation} \\
	\beta &=& \frac{(1-f_0)^2 \left[1-(1-v^2)f_0\right]\left[2r_0(f'_0)^2-f_0(3f'_0+r_0f''_0)\right]}{r_0^2(f'_0)^3} \quad .\label{eq:betar0relation}
\end{IEEEeqnarray} 
When integrating, we have%
\begin{IEEEeqnarray*}{rCl}
	\int_0^1 \frac{1}{\sqrt{\alpha z+\beta z^2}}&=& \frac{2\sinh^{-1}\left(\sqrt{\frac{\beta}{\alpha}}\right)}{\sqrt{\beta}}=\frac{2}{\sqrt{\beta}}\ln \left(\frac{\sqrt{\beta}+\sqrt{\alpha+\beta}}{\sqrt{\alpha}}\right)\to \frac{\ln \left[\frac{4\beta_c}{a\delta}\right]}{\sqrt{\beta_c}}
\end{IEEEeqnarray*} 
for small $\delta$, where we have substituted $\alpha=a \delta$ and $\beta=\beta_c$. Constants $a$ and $\beta_c$ can be obtained by first substituting $r_0=r_c(1+\delta)$ into the relations \eqref{eq:alphar0relation} and \eqref{eq:betar0relation} among $\alpha,~\beta$ and $r_0$, 
expanding the result respect to $\delta$ and then substituting the expression of $v$ and $f_c$ given by \eqref{eq:vfrcrule}
\begin{IEEEeqnarray}{rCl}
	 a&=& \frac{2f'_cr_c}{1-f_c}\beta_c \quad, \quad 
	 \beta_c = \frac{1}{2}(1-f_c)^2 \left[\frac{2}{f_c}-\frac{3f'_c+r_cf''_c}{r_c(f'_c)^2}\right] \quad.  \label{eq:abetaexpansion}
\end{IEEEeqnarray} 
When $\delta\to0$, $\alpha$ vanishes, which is the reason why the deflection angle diverges. Expanding $R_{\varphi}(z,r_0,v)$ around $z=0$ and $r_0\to r_c$ as well, we obtain the coefficients in the expansion~\eqref{eq:phi0divergencedelta} as
\begin{IEEEeqnarray}{rCl}
	\lambda_{\varphi} &=& \frac{R_\varphi(0,r_c,v)}{\sqrt{\beta_c}} \quad, \IEEEyesnumber\IEEEyessubnumber \\ 
	\zeta_{\varphi,D} &=& \lambda_{\varphi} \ln \left(\frac{4\beta_c}{a r_c}\right) = \lambda_{\varphi} \ln \left[\frac{2(1-f_c)}{f'_cr_c}\right] \quad, \IEEEyessubnumber\label{eq:defzetaD} \\ 
	\zeta_{\varphi,R} &=& \int_0^1\psi_{\varphi}(z,r_c,v)\dd z\quad,  \IEEEyessubnumber\label{eq:defzetaR}
\end{IEEEeqnarray} 
where $\psi_{\varphi}(z,r_0,v)$ is the remaining integrand in the $\varphi_0$ integral \eqref{eq:phi0integral} after subtracting the divergent term 
\begin{IEEEeqnarray}{rCl}
	\psi_{\varphi}(z,r_0,v) &=& \frac{R_\varphi(z,r_0,v)}{\sqrt{P(z,r_0,v)}}-\frac{R_\varphi(0,r_c,v)}{\sqrt{\alpha z+\beta z^2}} \quad. 
\end{IEEEeqnarray} 
After replacing $r_0$ by $r_c$ we obtain 
\begin{IEEEeqnarray}{rCl}
	\psi_{\varphi}(z,r_c,v) &=&  \frac{R_\varphi(z,r_c,v)}{\sqrt{P(z,r_c,v)}}-\frac{R_\varphi(0,r_c,v)}{\sqrt{\beta_c}z}\quad. 
\end{IEEEeqnarray} 
Substituting speed $v$ by its relation \eqref{eq:vfrcrule} with $r_c$, we see that $P(z,r_c,v)$ can be simplified to 
\begin{IEEEeqnarray}{rCl}
	P(z,r_c,v) &=& f_c-f(r)+\frac{r_cf(r)(r^2-r_c^2)f'_c}{2f_cr^2} \quad. 
\end{IEEEeqnarray}

\paragraph{The backward scattering.} For a massless particle, when it winds around the black hole one or more times (i.e., $\Delta\varphi=2\varphi_0-\pi=(2k-1)\pi$ with $\varphi_0=k\pi$ for $k=1,~2,~3,~\cdots$) and returns to the direction it came from, the cross-section can be approximated by the so-called \emph{glory approximation}~\cite{Zhang:1984vt,Dolan:2006vja}, which in our notation reads\footnote{For massless particles we have $v=1$, and we can't write $E=\frac{1}{\sqrt{1-v^2}}$ any more.}
\begin{IEEEeqnarray}{rCl}
	\frac{\dd\sigma}{\dd\Omega} &=&2\pi E \eta_{k\pi}^2\left|\frac{\dd b}{\dd\theta}\right|_{\theta=\pi}J_{2s}^2(E\eta_{k\pi}\sin\theta)  \quad. \label{eq:backcrosssection}
\end{IEEEeqnarray} 
In Eq.~\eqref{eq:backcrosssection}, $J_{2s}$ is the Bessel function and $\eta_{k\pi}$ is the corresponding impact parameter\footnote{Since $\eta=bv$, when $v=1$ we have $\eta=b$ is the impact parameter.} corresponds to $\varphi_0=k\pi$. Note that the glory usually appears in a strong lensing scenario, $\eta_{k\pi}$ can then be determined by the expansion~\eqref{eq:etar0relation}. Using the expansion~\eqref{eq:phi0divergencedelta} of $\varphi_0$, we can write $\delta$ as 
\begin{IEEEeqnarray*}{rCl}
	\delta &=& \exp \left(\frac{\zeta_{\varphi,D}+\zeta_{\varphi,R}-\varphi_0}{\lambda_\varphi}\right) \quad. 
\end{IEEEeqnarray*} 
Substituting into Eq.~\eqref{eq:etar0relation}, using the expression~\eqref{eq:r0etagammav1} and note that $\varphi_0=k\pi$ in the backward scattering we obtain
\begin{IEEEeqnarray}{rCl}
	\eta_{k\pi} &=&\eta_c\left(1+\gamma|_{v=1}\delta^2\right)= \eta_c+ \eta_c\left(\frac{1}{2}-\frac{\eta_c^2f''_c}{4}\right) \exp \left[\frac{2\left(\zeta_{\varphi,D}+\zeta_{\varphi,R}-k\pi\right)}{\lambda_\varphi}\right] \quad. 
\end{IEEEeqnarray} 
Therefore, once the constants $\lambda_{\varphi},~\zeta_{\varphi,D}$ and $\zeta_{\varphi,R}$ are known, the impact parameter for a given deflection angle can be determined easily in the \acrshort{sdl} scenario.
More generally, for a particle (possibly massive) with an arbitrary $\varphi_0$ in the strong deflection, we have 
\begin{IEEEeqnarray}{rCl}
	\eta_{\varphi_0} &=& \eta_c+ \gamma\eta_c\exp \left[\frac{2\left(\zeta_{\varphi,D}+\zeta_{\varphi,R}-\varphi_0\right)}{\lambda_\varphi}\right], \label{eq:etaphi0strong}
\end{IEEEeqnarray} 
where $\gamma$ is given by Eq.~\eqref{eq:r0etagamma}. Eq.~\eqref{eq:etaphi0strong} generalizes the results in Schwarzschild spacetime~\cite{Darwin:1959gfp, Dolan:2006vja} to more general spherically symmetric spacetimes.

\section{The strong deflection limit of the precession angle}
\label{sec:spinpreceession}
When a particle moves along a geodesic, its spin vector will be parallel transported~\cite{Pang:2024tco}. Let $S^\mu$ be the $4$-spin of the particle, then the parallel transport equation reads 
\begin{IEEEeqnarray}{rCl}
	\frac{\mathrm{D} S^\mu}{\dd\tau} &=& \frac{\dd S^\mu}{\dd\tau}+\Gamma^{\mu}_{\nu\lambda}\frac{\dd x^\lambda}{\dd\tau} \quad, \label{eq:Smuparallel}
\end{IEEEeqnarray} 
where $\mathrm{D}$ is the covariant differential operator, $\Gamma^{\mu}_{\nu\lambda}$ are Christoffel symbols, $\tau$ is the affine parameter along the geodesic, and $x^{\lambda}=(t,r,\theta,\varphi)$ represent coordinates.
After restricting to the equatorial plane, one can see that $S^\theta=0$ is a solution to the parallel transportation equation \eqref{eq:Smuparallel}\footnote{In fact, the $S^\theta$ equation decouples from the others, and a spin vector that is perpendicular to the equatorial plane will remain so during the parallel transport~\cite{Pang:2024tco}. Similarly, when focusing on the precession of a particle's spin that lies in the equatorial plane, one can safely set $S^\theta=0$.}. Furthermore, the $S^t$ component can be eliminated by using the orthogonal relation $S^\mu U_\mu = 0$
with $4$-velocity $U_\mu=\frac{\dd x_\mu}{\dd \tau}$~\cite{Weinberg:1972kfs}.
As a result, we can focus on the $S^r$ and $S^\varphi$ components of the spin vector. After change the variable from $\tau$ to $\varphi$, the parallel transport equation \eqref{eq:Smuparallel} can be rewritten as~\cite{Pang:2024tco}
\begin{IEEEeqnarray}{rCl}
	 && \frac{\dd}{\dd\varphi} S^r(\varphi)-\left[f(r)-\frac{r}{2}\frac{\dd f}{\dd r}\right]\bar{S}^\varphi(\varphi) =0  \quad,  \IEEEyesnumber \IEEEyessubnumber \label{eq:srpphi}\\ 
	 &&\frac{\dd }{\dd\varphi}\bar{S}^\varphi(\varphi)+S^r(\varphi)=0 \quad , \IEEEyessubnumber \label{eq:sphipphi}
\end{IEEEeqnarray} 
where we have defined $\bar{S}^\varphi=r S^\varphi$ for convenience. %
Finally, we can take a further derivative respect to $\varphi$ on the both sides of Eq.~\eqref{eq:sphipphi} and then substitute $\frac{\dd}{\dd\varphi}S^r$ from Eq.~\eqref{eq:srpphi} to obtain a second-order differential equation for $\bar{S}^\varphi$~\cite{Pang:2024tco}
\begin{IEEEeqnarray}{rCl}
	\frac{\dd^2}{\dd\varphi^2}\bar{S}^\varphi + \left[f(r)-\frac{r}{2}\frac{\dd f}{\dd r}\right]\bar{S}^\varphi(\varphi)&=&0  \quad. \label{eq:sphippphi}
\end{IEEEeqnarray} 

\paragraph{The precession angle.} By introducing the precession angle  $\chi$ that satisfies~\cite{Dolan:2006vja,Pang:2024tco}
\begin{IEEEeqnarray}{rCl}
	\frac{\dd\chi}{\dd\varphi} &=& E \left[1+\frac{L^2}{r(\varphi)^2}\right]^{-1} = \frac{\sqrt{1-v^2}r(\varphi)^2}{(1-v^2)r(\varphi)^2+\eta^2}\quad, \label{eq:chipphi}
\end{IEEEeqnarray} 
we can rewrite the $\bar{S}^\varphi$ equation \eqref{eq:sphippphi} in a simpler form 
\begin{IEEEeqnarray}{rCl}
	\frac{\dd^2B}{\dd\chi^2}+B &=& 0 \quad, \label{eq:Bppchi}
\end{IEEEeqnarray} 
where $B$ is defined as 
\begin{IEEEeqnarray}{rCl}
	B[\chi(\varphi)] &=& \frac{r(\varphi)\bar{S}^{\varphi}(\varphi)}{\sqrt{r(\varphi)^2+L^2}} = \frac{\sqrt{1-v^2}r(\varphi)\bar{S}^{\varphi}(\varphi)}{\sqrt{(1-v^2)r(\varphi)^2+\eta^2}} \quad. 
\end{IEEEeqnarray} 

Solving the $B$ equation \eqref{eq:Bppchi} is straightforward. In fact, with the initial condition $S^r(0)=0$ and $S^{\varphi}(0)=1$, the precession Eqs.~\eqref{eq:srpphi} and \eqref{eq:sphipphi} can be solved in terms of the precession angle $\chi$ as
\begin{IEEEeqnarray}{rCl}
	\bar{S}^\varphi(\varphi) &=&\frac{\sqrt{(1-v^2)r(\varphi)^2+\eta^2}}{\sqrt{1-v^2}r(\varphi)} \cos[\chi(\varphi)]\quad,\IEEEyesnumber\IEEEyessubnumber \label{eq:barSphichisol} \\ 
	S^r(\varphi) &=& \frac{r(\varphi)}{\sqrt{(1-v^2)r(\varphi)^2+\eta^2}}\sin[\chi(\varphi)] \nonumber \\ 
				 &&+ \frac{\eta}{\sqrt{1-v^2}}\sqrt{\frac{1-(1-v^2)V_{\text{eff}}[r(\varphi)]}{(1-v^2)r(\varphi)^2+\eta^2}}\cos[\chi(\varphi)]  \quad. \IEEEyessubnumber  \label{eq:Srchisol}
\end{IEEEeqnarray} 
Therefore, the precession angle $\chi$ characterizes the precession of the spin vector in the equatorial plane, as shown in Fig.~\ref{fig:anglephichixi}. 

Let $\vec{S}_i$ and $\vec{S}_f$ be the initial and final spin vectors of the particle, respectively. We can define $\xi$ to be an angle  between them, such that 
\begin{IEEEeqnarray}{rCl}
	\cos\xi &=& \frac{\vec{S}_i\cdot\vec{S}_f}{\left|\vec{S}_i\right|\left|\vec{S}_f\right|} \quad. 
\end{IEEEeqnarray} 
The relation among $\xi,~\varphi_0$ and $\chi_0=\chi(r_0)$ can be given by the following equation~\cite{Pang:2024tco}
\begin{IEEEeqnarray}{rCl}
	\xi &=& 2\chi_0-2\varphi_0+2\pi \quad.  \label{eq:defxichi0phi0}
\end{IEEEeqnarray} 

\paragraph{The integral form.} To see the divergent structure of the precession angle $\chi$, we can first change the variable in the definition~\eqref{eq:chipphi} from $\varphi$ to $z$ (the variable we introduced before through equation \eqref{eq:varzdef}) to obtain
\begin{IEEEeqnarray*}{rCl}
	\frac{\dd\chi}{\dd z} &=& \frac{\dd \chi}{\dd\varphi}\frac{\dd\varphi}{\dd z}=\frac{\sqrt{1-v^2}r(z)^2}{(1-v^2)r(z)^2+\eta^2}\frac{R_\varphi(z,r_0,v)}{\sqrt{P(z,r_0,v)}} \quad. 
\end{IEEEeqnarray*} 
Therefore, we can write 
\begin{IEEEeqnarray}{rCl}
	\frac{\dd \chi}{\dd z} &=& \frac{R_{\chi}(z,r_0,v)}{\sqrt{P(z,r_0,v)}} \quad, \label{eq:chipzRchiP}
\end{IEEEeqnarray} 
where we have defined the function $R_{\chi}(z,r_0,v)$ as
\begin{IEEEeqnarray}{rCl}
	R_{\chi}(z,r_0,v) &=& \frac{\sqrt{1-v^2}r(z)^2}{(1-v^2)r(z)^2+\eta^2}R_\varphi(z,r_0,v) \quad, \nonumber \\
					  &=& \frac{\sqrt{1-v^2}r(z)^2f_0}{(1-v^2)r(z)^2f_0+r_0^2 \left[1-(1-v^2)f_0\right]} R_\varphi(z,r_0,v)\quad,  \label{eq:defRchi}
\end{IEEEeqnarray} 
where  the relation~\eqref{eq:etar0relation} is used to eliminate $\eta$.
And as we have emphasized before, $r(z)$ is a function of $z$ instead of an independent variable, therefore, same as $R_{\varphi}(z,r_0,v)$, the new function $R_{\chi}(z,r_0,v)$ is also a function of $z$ instead of $r$. 

Same as $\varphi_0$ in equation \eqref{eq:phi0integral}, the precession angle $\chi_0$ can also be written as an integral based on the differential Eq.~\eqref{eq:chipzRchiP}
\begin{IEEEeqnarray}{rCl}
	\chi_0 &=& \int_0^1\dd z \frac{R_{\chi}(z,r_0,v)}{\sqrt{P(z,r_0,v)}}\quad.  \label{eq:chi0integral}
\end{IEEEeqnarray} 
Since the denominator of \eqref{eq:chi0integral} and \eqref{eq:phi0integral} are the same, we would expect that $\chi_0$ also diverges logarithmically as $r_0\to r_c$.

\paragraph{The strong deflection limit.}
By the same procedure as for the expansion of $\varphi$, we can expand $\chi$ as 
\begin{IEEEeqnarray}{rCl}
	\chi_0 &=& -\lambda_{\chi}\ln \delta +\zeta_{\chi,D}+\zeta_{\chi,R} \quad, 
\end{IEEEeqnarray} 
where 
\begin{IEEEeqnarray}{rCl}
	\lambda_{\chi} &=& \frac{R_\chi(0,r_c,v)}{\sqrt{\beta_c}} \quad,\IEEEyesnumber \IEEEyessubnumber \label{eq:deflambdachi}\\ 
	\zeta_{\chi,D} &=& \lambda_{\chi}\ln \left(\frac{4\beta_c}{a r_c}\right) = \lambda_{\chi}\ln \left[\frac{2(1-f_c)}{f'_cr_c}\right] \quad, \IEEEyessubnumber\label{eq:defzetachid}\\ 
	\zeta_{\chi,R} &=& \int_0^1\psi_{\chi}(z,r_c,v)\dd z \quad. \IEEEyessubnumber\label{eq:defzetachir}
\end{IEEEeqnarray} 
For the precession angle $\chi$ we have defined $\psi_{\chi}$ as \begin{IEEEeqnarray}{rCl}
	\psi_{\chi}(z,r_0,v) &=& \frac{R_\chi(z,r_0,v)}{\sqrt{P(z,r_0,v)}}-\frac{R_\chi(0,r_c,v)}{\sqrt{\beta_c }z} \quad.  \label{eq:defetachizr}
\end{IEEEeqnarray} 

One can show that $\psi_{\chi}(0,r_c,v)$ is finite and that the integral $\zeta_{\chi,R}$ converges, just as $\zeta_{\varphi,R}$ does. 

From the relation~\eqref{eq:defRchi} between $R_{\chi}$ and $R_{\varphi}$ we obtain
\begin{IEEEeqnarray}{rCl}
    \lambda_{\chi} = \sqrt{1-v^2}f_c\,\lambda_{\varphi} \quad,\quad
    \zeta_{\chi,D} = \sqrt{1-v^2}f_c\,\zeta_{\varphi,D} \quad, \label{eq:lambdachiphirelation}
\end{IEEEeqnarray}
which yields
\begin{IEEEeqnarray*}{rCl}
    -\lambda_{\chi}\ln\delta + \zeta_{\chi,D}
    &=& \sqrt{1-v^2}f_c\bigl(-\lambda_{\varphi}\ln\delta + \zeta_{\varphi,D}\bigr)
     =  \sqrt{1-v^2}f_c\bigl(\varphi_0 - \zeta_{\varphi,R}\bigr) \quad,
\end{IEEEeqnarray*}
and therefore $\chi_0$ can be written as
\begin{IEEEeqnarray}{rCl}
    \boxed{\chi_0 = \sqrt{1-v^2}f_c\,\varphi_0 - \sqrt{1-v^2}f_c\,\zeta_{\varphi,R} + \zeta_{\chi,R}} \quad. \label{eq:chi0phi0relation}
\end{IEEEeqnarray}
It's our main result, which establishes a direct relation between the precession angle $\chi_0$ and the deflection angle $\varphi_0$ in the \acrshort{sdl}. In particular, for backward scattering $\varphi_0 = k\pi$ (with $k$ an integer) is known a priori; thus $\chi_0$ is fully determined once $f_c$, $\zeta_{\varphi,R}$ and $\zeta_{\chi,R}$ are evaluated. 

Although we previously examined spin precession in Schwarzschild and \acrshort{rn} spacetimes~\cite{Pang:2024tco}, the remarkably compact relation~\eqref{eq:chi0phi0relation} emerges naturally only within the \acrshort{sdl} formalism. Articulating this connection in general spherically symmetric spacetimes therefore constitutes the principal contribution of the current work.

\paragraph{Massless particles.} For massless particles, the spin vector (lying in the equatorial plane) is always reversed after a backward scattering, resulting in $\cos\xi=-1$~\cite{Dolan:2006vja,Pang:2024tco}. Substituting $v=1$ into the definition~\eqref{eq:defxichi0phi0} of $\xi$ and the expansion~\eqref{eq:chi0phi0relation} of $\chi_0$ yields
\begin{IEEEeqnarray}{rCl}
    \xi &=& 2\zeta_{\chi,R} - 2\varphi_0 + 2\pi \quad.
\end{IEEEeqnarray}
Then $\cos\xi=-1$ and $\varphi_0=k\pi$
implies $\zeta_{\chi,R}=\frac{\pi}{2}$. Recalling that, for $v=1$, $\zeta_{\chi,R}$ is defined by the integral of $\psi_{\chi}(z,r_0,1)$ in Eq.~\eqref{eq:defetachizr}, we conclude that
\begin{IEEEeqnarray}{rCl}
    \lim_{v\to 1}\int_0^1\dd z \left[\frac{R_\chi(z,r_c,v)}{\sqrt{P(z,r_c,v)}}-\frac{R_\chi(0,r_c,v)}{\sqrt{\beta_c}\,z}\right] &=& \frac{\pi}{2}
\end{IEEEeqnarray}
for any asymptotically flat blackening factor $f(r)$\footnote{Recall that the $f(r)$ enters through $R_\chi(z,r_0,v)$, defined in Eq.~\eqref{eq:defRchi}, with $r$ understood as a function of $z$ via the change of variable~\eqref{eq:varzdef}.}.

\section{Examples}
\label{sec:examples}
\subsection{Schwarzschild spacetime}
\label{sec:exaSch}
In Schwarzschild spacetime, the blackening factor $f(r)$ reads
\begin{IEEEeqnarray}{rCl}
    f(r) &=& 1-\frac{2M}{r} = 1-\frac{1}{r} \quad, \label{eq:metricsch}
\end{IEEEeqnarray}
where we have set $M=1/2$ so that the horizon radius is $r_h=1$. 
Solving the two conditions~\eqref{eq:rcdefeq} yields the critical radius~\cite{Bozza:2002zj,Feleppa:2024kio}
\begin{IEEEeqnarray}{rCl}
    r_c &=& \frac{4v^2+\sqrt{8v^2+1}-1}{4v^2}
         =  \frac{3(x+1)}{3x+1} \label{eq:rcrelationSchx} \quad ,\quad 
    x  = \frac{\sqrt{1+8v^2}}{3} \quad ,
\end{IEEEeqnarray}
where the auxiliary parameter $x\in[1/3,1]$ is a monotonically increasing function of the particle speed $v\in[0,1]$.

In terms of $x$, we see that 
\begin{IEEEeqnarray}{rCl}
    f_c &=& f(r_c)
         =  \frac{2}{3(x+1)} \quad. \label{eq:fcrelationSchx}
\end{IEEEeqnarray}

\begin{figure}[htpb]
    \centering
    \includegraphics[width=0.85\textwidth]{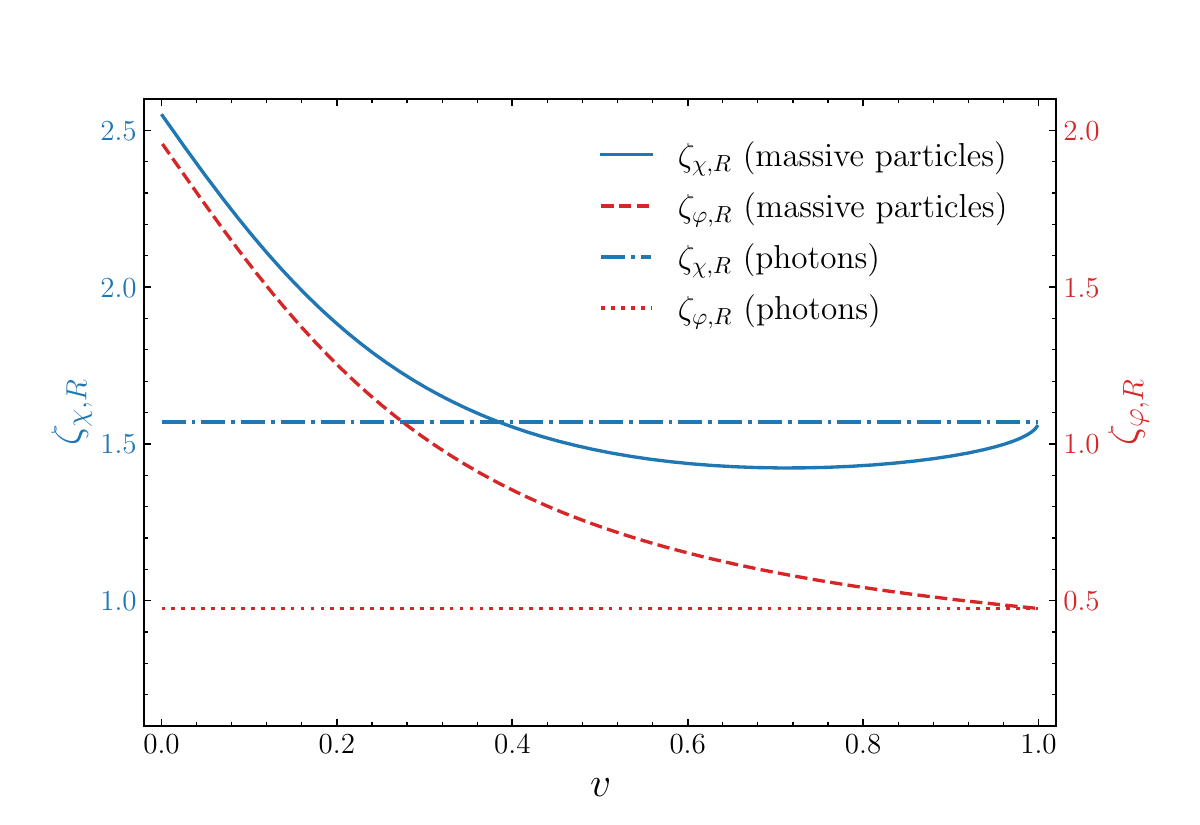}
\caption{Regular part of \acrshort{sdl} coefficients as functions of the particle speed $v$ in the Schwarzschild spacetime.
    The blue solid and dash-dotted curves show $\zeta_{\chi,R}$ [Eq.~\eqref{eq:zetachiRSch}] for massive and massless particles, respectively, scaled by the left axis.
    The red dashed and dotted curves show $\zeta_{\varphi,R}$ [Eq.~\eqref{eq:zetaphiRSch}] for massive and massless particles, respectively, scaled by the right axis.
    As expected, $\zeta_{\chi,R}\to\pi/2$ as $v\to1$, consistent with the universal spin flip for massless particles.}
    \label{fig:shwall}
\end{figure}

\paragraph{The deflection angle.}
The \acrshort{sdl} expansion of the deflection angle $\varphi_0$ in the Schwarzschild spacetime is well established~\cite{Tsupko:2014wza,Liu:2015zou,Feleppa:2024kio}.
Adopting the notation defined in Eqs.~\eqref{eq:rcrelationSchx} and~\eqref{eq:fcrelationSchx}, and substituting the metric coefficients from Eq.~\eqref{eq:metricsch} into~\eqref{eq:abetaexpansion}, we obtain
\begin{IEEEeqnarray}{rCl}
    a &=& 2\beta_c \quad,\quad \beta_c = \frac{x(3x+1)}{2(x+1)} \quad.
    \label{eq:abetaexpansionSchx}
\end{IEEEeqnarray}
The coefficients governing the logarithmic divergence of $\varphi_0$, given by the expansion in Eq.~\eqref{eq:phi0divergencedelta}, take the following explicit forms
\begin{IEEEeqnarray}{rCl}
   \lambda_{\varphi}& = &2 \sqrt{\frac{1+x}{2x}} \quad ,\IEEEyesnumber\IEEEyessubnumber\label{eq:lambdaphiSch} \\
  \zeta_{\varphi,D} &=& \lambda_{\varphi}\ln (2) \quad , \IEEEyessubnumber\label{eq:zetaphiDSch}\\
  \zeta_{\varphi,R}& =& 2\lambda_{\varphi} \left[ \ln(12) - \ln\left(\sqrt{18-\frac{6}{x}}+6\right) \right] \quad. \IEEEyessubnumber\label{eq:zetaphiRSch}
\end{IEEEeqnarray}
Since the dimensionless parameter $x$ increases monotonically with the particle speed $v$ (ranging from $0$ to $1$), both $\lambda_{\varphi}$ and $\zeta_{\varphi,R}$ exhibit a monotonic decrease as $v$ increases.
The latter behavior is shown in Fig.~\ref{fig:shwall}, where $\zeta_{\varphi,R}$ is plotted as a function of $v$.

\paragraph{The precession angle.}
The \acrshort{sdl} expansion for the precession angle $\chi_0$ is derived analogously to the deflection angle.
Substituting the expressions for $f_c$, $\lambda_{\varphi}$ and $\zeta_{\varphi,D}$ from Eqs.~\eqref{eq:fcrelationSchx}, \eqref{eq:lambdaphiSch} and \eqref{eq:zetaphiDSch} into the relation~\eqref{eq:lambdachiphirelation} yields
\begin{IEEEeqnarray}{rCl}
    \lambda_\chi &=& \frac{1}{2} \sqrt{\frac{1}{x}-1} \quad, \IEEEyesnumber\IEEEyessubnumber
    \label{eq:lambdachiSch} \\
    \zeta_{\chi,D}&=&\lambda_{\chi}\ln 2 \quad . \IEEEyessubnumber \label{eq:zetachiDSch}
\end{IEEEeqnarray}

In the Schwarzschild spacetime, the integral~\eqref{eq:defzetachir} defining  $\zeta_{\chi,R}$ admits an analytic evaluation as the following 
\begin{IEEEeqnarray}{rCl}
\zeta_{\chi,R} &=&
\tan^{-1}\!\left( \frac{3\sqrt{x(1-x)}}{1-3x} \right)
- \tan^{-1}\!\left( \sqrt{\frac{2-6x}{3x-3}} \right) + \pi \nonumber \\
&& + \lambda_{\chi} \left[\ln (12)-\ln \left(\sqrt{18-\frac{6}{x}}+6\right)\right] \quad . \IEEEyessubnumber\label{eq:zetachiRSch}
\end{IEEEeqnarray}

As illustrated in Fig.~\ref{fig:shwall}, in contrast to the monotonic decrease behavior of $\zeta_{\varphi,R}$, for the precession angle, $\zeta_{\chi,R}$ exhibits a non-monotonic dependence on particle speed $v$, it initially decreases before rising as $v$ increases.
Furthermore, in the ultra-relativistic limit ($v\to 1$), $\zeta_{\chi,R}$ asymptotes to $\pi/2$, consistent with the previously established massless limit.

\begin{figure}[htpb]
    \centering
    \includegraphics[width=0.85\textwidth]{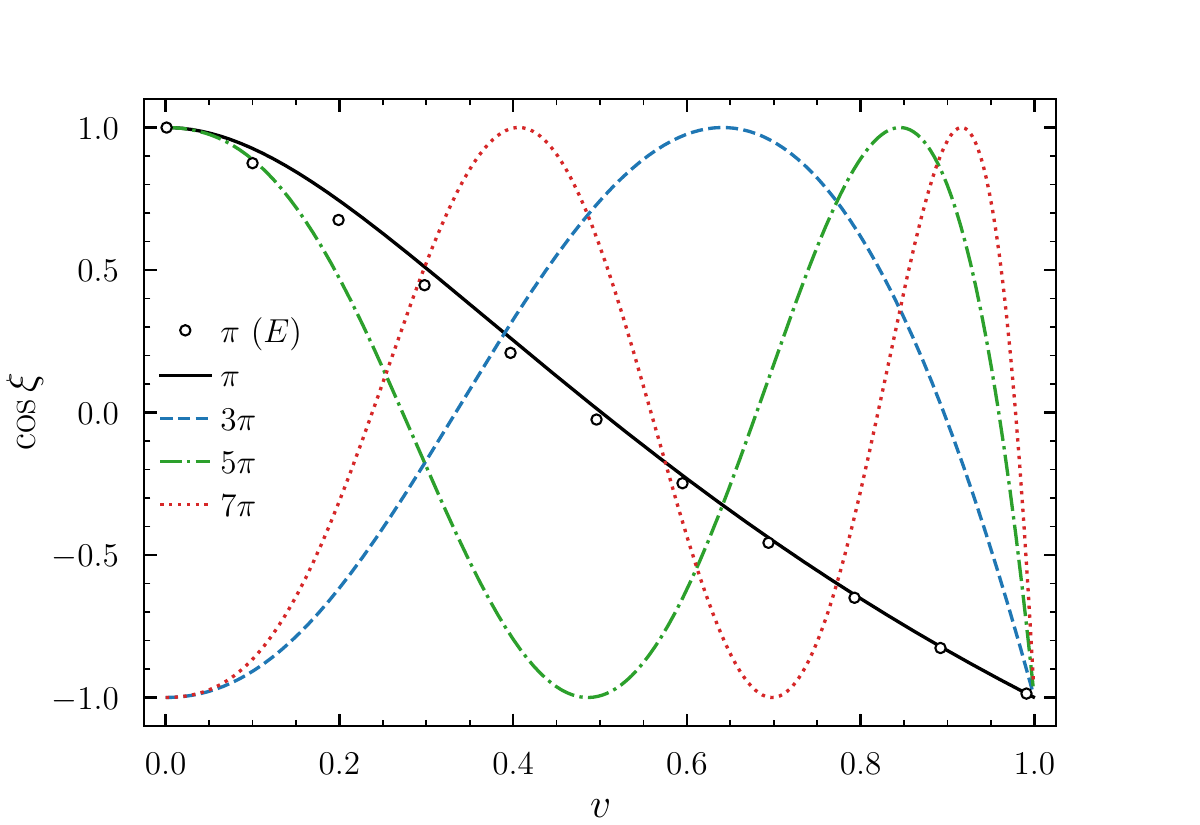}
    \caption{$\cos\xi$ as a function of the particle speed $v$ for backward scattering in the Schwarzschild spacetime.
    The black solid, blue dashed, green dash-dotted, and red dotted curves represent the \acrshort{sdl} expansion for deflection angle $\Delta\varphi = \pi, 3\pi, 5\pi,$ and $7\pi$, respectively.
    The black circles correspond to the exact numerical result for $\Delta\varphi = \pi$~\cite{Pang:2024tco}.
    The deviation between the expansion and the exact result at $\Delta\varphi=\pi$ for small $v$ reflects the breakdown of the \acrshort{sdl} expansion~\eqref{eq:chi0phi0relation} away from the strong-deflection limit, as explained in the main text.}
    \label{fig:cosxiSch}
\end{figure}

Substituting $f_c$, $\zeta_{\varphi,R}$, and $\zeta_{\chi,R}$ from Eqs.~\eqref{eq:fcrelationSchx}, \eqref{eq:zetaphiRSch}, and~\eqref{eq:zetachiRSch} into the master relation~\eqref{eq:chi0phi0relation} between $\chi_0$ and $\varphi_0$, and subsequently into the definition~\eqref{eq:defxichi0phi0}, we obtain the angle $\xi$ between the initial and final spin vectors $\vec{S}_i$ and $\vec{S}_f$.
Fig.~\ref{fig:cosxiSch} depicts $\cos\xi$ as a function of $v$ for several winding numbers.
As expected, $\cos\xi \to -1$ in the limit $v\to 1$, corresponding to a complete spin flip.
For comparison, we also plot the exact result for $\Delta\varphi = \pi$~\cite{Pang:2024tco}.
The slight discrepancy observed at lower velocities arises because $\Delta\varphi = \pi$ lies outside the deep \acrshort{sdl} regime for non-relativistic particles; in other words, the expansion parameter $\delta = (r_0 - r_c)/r_c$ is not sufficiently small to ensure the accuracy of the asymptotic expansion.
Results for higher winding numbers ($\Delta\varphi= 3\pi, 5\pi, 7\pi$) are nearly indistinguishable from the expansion result and are omitted for clarity (cf. Fig.~3 in \cite{Pang:2024tco}).

\subsection{\acrlong{rn} spacetime}
\label{sec:exaRN}
In \acrshort{rn} spacetime, the blackening factor $f(r)$ reads
\begin{IEEEeqnarray}{rCl}
	f(r) &=& 1-\frac{2M}{r}+\frac{Q^2}{r^2} 
	      =1-\frac{1}{r}+\frac{h^2}{4r^2} \quad, \label{eq:RNmetric}
\end{IEEEeqnarray}
where $Q$ is the charge of the black hole. As in the Schwarzschild case, we set the black hole mass to $M=1/2$ and introduce the dimensionless charge parameter $h\equiv Q/M$ (with $0\leq h\leq 1$ to avoid a naked singularity).

Unlike the Schwarzschild case, closed-form analytic expressions for $\varphi_0$ and $\chi_0$ in the \acrshort{rn} spacetime are hard to obtain in this \acrshort{sdl} formalism. We therefore adopt a perturbative approach, treating $h$ as a small expansion parameter and computing all \acrshort{sdl} coefficients to order $h^2$~\cite{Feleppa:2024kio}.
The critical radius $r_c$ then reads
\begin{IEEEeqnarray}{rCl}
	r_c &=& \frac{3(x+1)}{3 x+1}-\frac{h^2}{4} \left(\frac{1}{3 x}+1\right) \quad.	 \label{eq:rcrelationRN}
\end{IEEEeqnarray}
where the first term recovers the Schwarzschild result~\eqref{eq:rcrelationSchx} and the $h^2$ correction encodes the charge dependence.

\paragraph{The deflection angle.}
The \acrshort{sdl} expansion of $\varphi_0$ in the \acrshort{rn} spacetime was recently obtained in~\cite{Feleppa:2024kio}.
Adopting the notation of Sec.~\ref{sec:deflectionangle} and substituting the \acrshort{rn} metric~\eqref{eq:RNmetric} into the general expressions~\eqref{eq:abetaexpansion}, we expand to order $h^2$ and obtain
\begin{IEEEeqnarray}{rCl}
	a &=& 2\beta_c\quad,\quad \beta_c=\frac{x (3 x+1)}{2(x+1)}+\frac{\left(3 x^2+1\right) (3 x+1)^2}{144 x (x+1)^2} h^2 \quad. \label{eq:abetaexpansionRN}
\end{IEEEeqnarray}
The coefficients governing the logarithmic divergence of $\varphi_0$ [Eq.~\eqref{eq:phi0divergencedelta}] take the following forms
\begin{IEEEeqnarray}{rCl}
\lambda_\varphi &=& 
\lambda_{\varphi,0}+\frac{\left(18 x^3+15 x^2-1\right) }{144 \sqrt{2} \sqrt{x^5 (x+1)}}h^2 \quad , \IEEEyesnumber\IEEEyessubnumber
\label{eq:lambdaphiRN} 
\\
\zeta_{\varphi,D} &=&
\zeta_{\varphi,D,0}+\frac{(3 x+1)^{3/2} \left[x^2 (12+\ln (64))+x \ln (8)-\ln (2)\right] }{144 \sqrt{2} x^2\sqrt{ \left(3 x^2+4 x+1\right)}} h^2\quad,	\IEEEyessubnumber
\label{eq:zetaphidRN} \\
\zeta_{\varphi,R} &=&
\zeta_{\varphi,R,0} + \zeta_{\varphi,R,1}\, h^2 \quad,\IEEEyessubnumber
\label{eq:zetaphirRNexpexp}
\end{IEEEeqnarray}
where $\lambda_{\varphi,0},~\zeta_{\varphi,D,0}$ and $\zeta_{\varphi,R,0}$ are the Schwarzschild results~\eqref{eq:lambdaphiSch}--\eqref{eq:zetaphiRSch}, and $\zeta_{\varphi,R,1}$ can be integrated analytically to give
\begin{IEEEeqnarray}{rCl}
    \zeta_{\varphi,R,1} &=&\frac{\left(18 x^3+15 x^2-1\right) \left[\ln (12)-\ln \left(\sqrt{18-\frac{6}{x}}+6\right)\right]}{72 \sqrt{2} \sqrt{x^5 (x+1)}} \nonumber \\
    &&- \frac{2 \sqrt{\frac{2}{x}+2} \left(54 x^3+9 x^2+2 x-1\right)-4 \sqrt{3} \sqrt{\frac{x+1}{3 x-1}} \left(18 x^3-3 x^2+2 x-1\right)}{288 x^2 (x+1)} \quad . \nonumber \\
    \label{eq:zetaphirRN}
\end{IEEEeqnarray}
In the massless limit $x\to 1$, we see that $\zeta_{\varphi,R,1}=\frac{1}{9}\bigl\{-4 + \sqrt{3} + \ln\bigl[-6(-2+\sqrt{3})\bigr]\bigr\}$, which is consistent with the previous result~\cite{Bozza:2002zj}.

\paragraph{The precession angle.}
The \acrshort{sdl} expansion for the precession angle $\chi_0$ can be derived analogously to the deflection angle.
Substituting the \acrshort{rn} metric into the general expressions~\eqref{eq:deflambdachi}--\eqref{eq:defzetachir} and expanding to order $h^2$ yields
\begin{IEEEeqnarray}{rCl}
\lambda_\chi &=&
\lambda_{\chi,0}+\frac{\sqrt{1-x^2}\bigl(3 x^2-2 x-1\bigr) }{288 x^{5/2} (x+1)^{3/2}} h^2\quad, \IEEEyesnumber\IEEEyessubnumber
\label{eq:lambdachiRN} \\
\zeta_{\chi,D} &=& \zeta_{\chi,D,0}+\frac{(3 x+1)\sqrt{1-x^2}\bigl[12 x^2 + x\ln (2) - \ln (2)\bigr]}{288 x^{5/2} (x+1)^{3/2}}h^2 \quad , \IEEEyessubnumber\label{eq:zetachidRN} \\
\zeta_{\chi,R} &=& \zeta_{\chi,R,0} + \zeta_{\chi,R,1} h^2 \quad, \IEEEyessubnumber
\label{eq:zetachirRN}
\end{IEEEeqnarray}
where $\lambda_{\chi,0},~\zeta_{\chi,D,0}$ and $\zeta_{\chi,R,0}$ are the Schwarzschild results~\eqref{eq:lambdachiSch}--\eqref{eq:zetachiRSch}, and $\zeta_{\chi,R,1}$ reads 
\begin{IEEEeqnarray}{rCl}
\zeta_{\chi,R, 1} &=&
\frac{\sqrt{1-x} \left(3 x^2-2 x-1\right) 
\left[\ln (12)-\ln \left(\sqrt{18-\frac{6}{x}}+6\right)\right]}{144 x^{5/2} (x+1)} \nonumber \\
&& -\,\frac{\sqrt{1-x} (3 x+1) \left[x (3 x+2)-\sqrt{6} \sqrt{x (3 x-1)}-1\right]}{144 x^{5/2} (x+1)} \quad .
\label{eq:zetachir1RN}
\end{IEEEeqnarray}

\begin{figure}[htpb]
	\centering
	\includegraphics[width=0.85\textwidth]{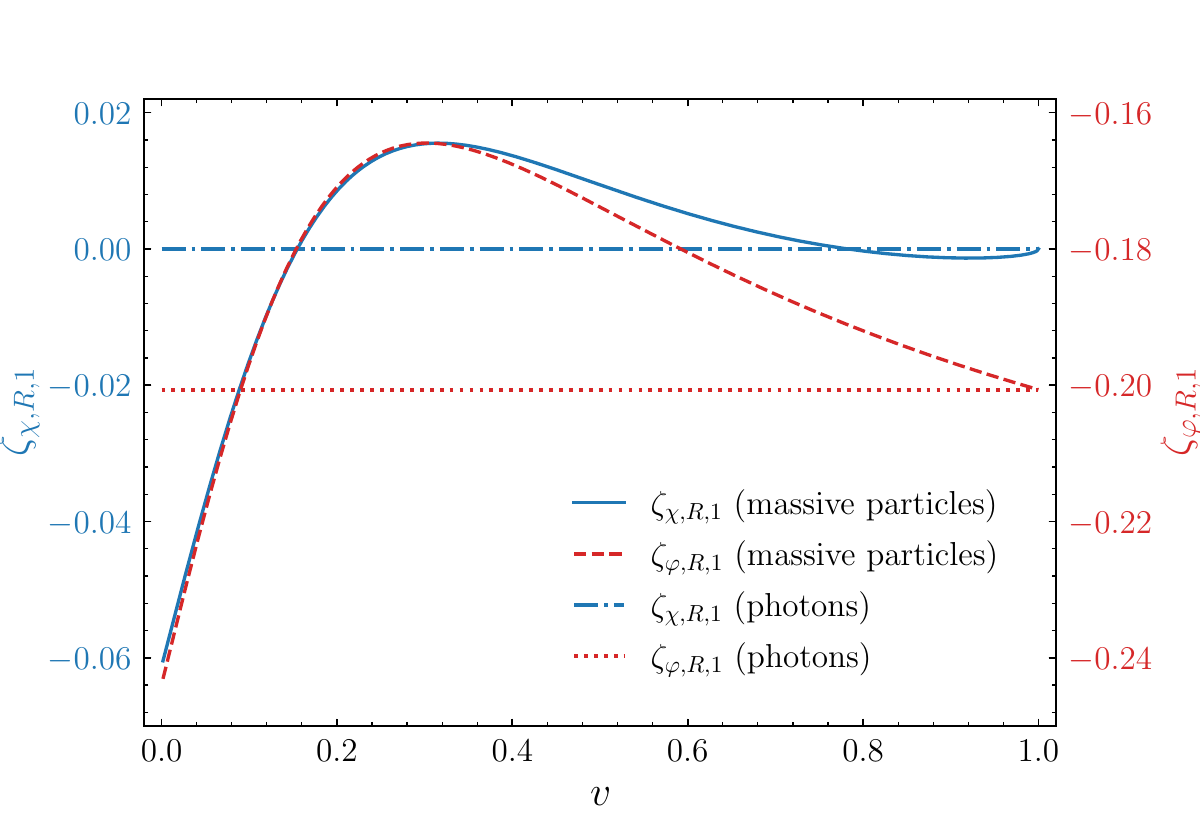}
	\caption{First-order charge corrections to the regular part of \acrshort{sdl} coefficients as functions of the particle speed $v$ in the \acrshort{rn} spacetime.
    The line styles follow the same convention as in Fig.~\ref{fig:shwall}: blue curves (left axis) represent $\zeta_{\chi,R,1}$ [Eq.~\eqref{eq:zetachir1RN}], and red curves (right axis) represent $\zeta_{\varphi,R,1}$ [Eq.~\eqref{eq:zetaphirRN}].
    Note that $\zeta_{\chi,R,1}\to0$ as $v\to1$, confirming that the charge has no effect on spin precession in the massless limit.}
	\label{fig:zatachirrn}
\end{figure}

As shown in Fig.~\ref{fig:zatachirrn}, $\zeta_{\chi,R,1}$ vanishes in the massless limit $v\to 1$, demonstrating that the black hole charge does not affect the spin precession of massless particles, at least to $Q^2$ order. This is consistent with the general expectation we have mentioned in Sec.~\ref{sec:spinpreceession}, i.e., $\cos\xi=-1$ (i.e., a complete spin flip) for massless particles in any asymptotically flat spacetime, independent of the metric function $f(r)$. For massive particles ($v<1$), however, the charge introduces a non-trivial correction that distinguishes the \acrshort{rn} case from Schwarzschild one.

\begin{figure}[htpb]
	\centering
	\includegraphics[width=0.85\textwidth]{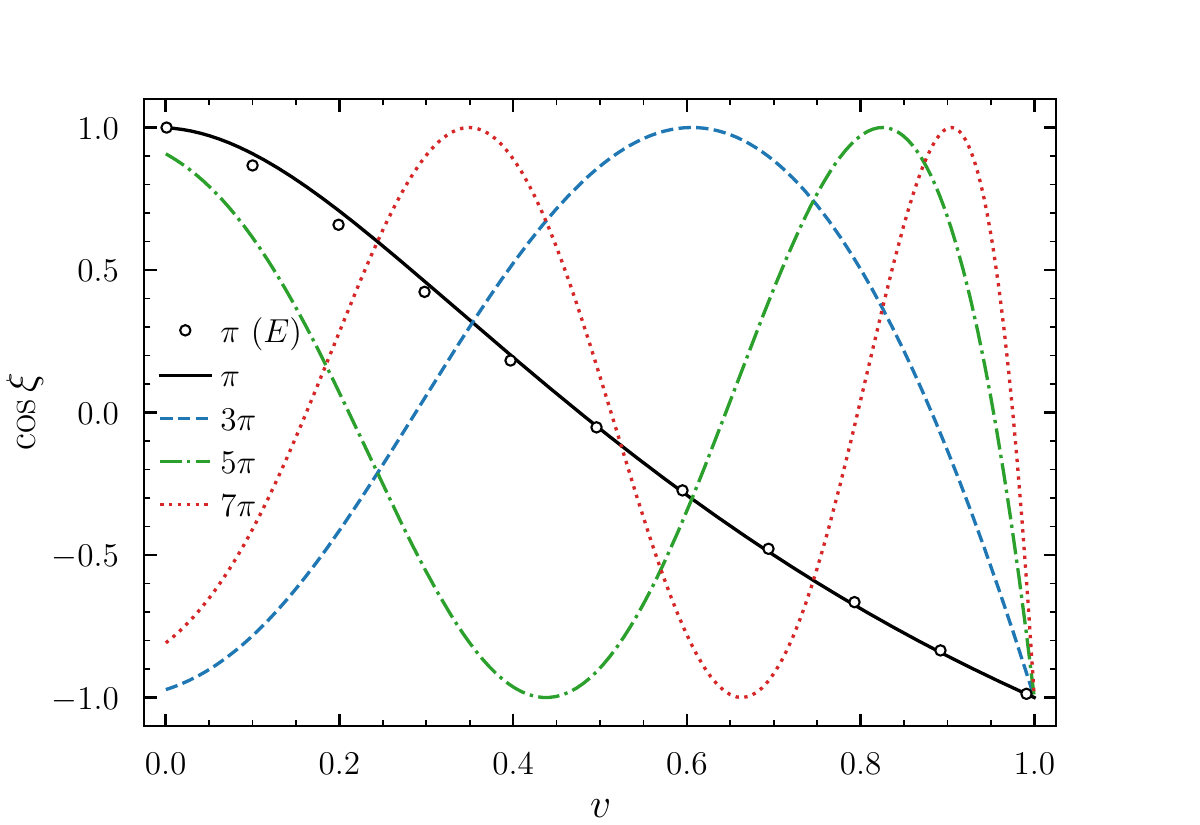}
	\caption{$\cos\xi$ as a function of the particle speed $v$ for backward scattering in the \acrshort{rn} spacetime with $h = 0.6$.
    The line styles follow the same convention as in Fig.~\ref{fig:cosxiSch}.
    The black circles correspond to the exact result for $\Delta\varphi = \pi$~\cite{Pang:2024tco}.
    The deviation between the expansion and the exact result at $\Delta\varphi=\pi$ for small $v$ reflects the same breakdown of the \acrshort{sdl} expansion already discussed for the Schwarzschild case (cf.\ Fig.~\ref{fig:cosxiSch}).}
	\label{fig:cosxirn}
\end{figure}

Substituting the \acrshort{sdl} coefficients into the master relation~\eqref{eq:chi0phi0relation} between $\chi_0$ and $\varphi_0$, and subsequently into the definition~\eqref{eq:defxichi0phi0} of $\xi$, we obtain the angle between the initial and final spin vectors, as shown in Fig.~\ref{fig:cosxirn}. 
Similar to the Schwarzschild case, we see that $\cos\xi\to-1$ as $v\to1$, confirming the universal spin flip for massless particles. The small deviation from the exact result at low $v$ for $\Delta\varphi=\pi$ is a consequence of the finite expansion parameter $\delta$ and was already discussed in the Schwarzschild context.

\begin{figure}[htpb]
	\centering
	\includegraphics[width=0.85\textwidth]{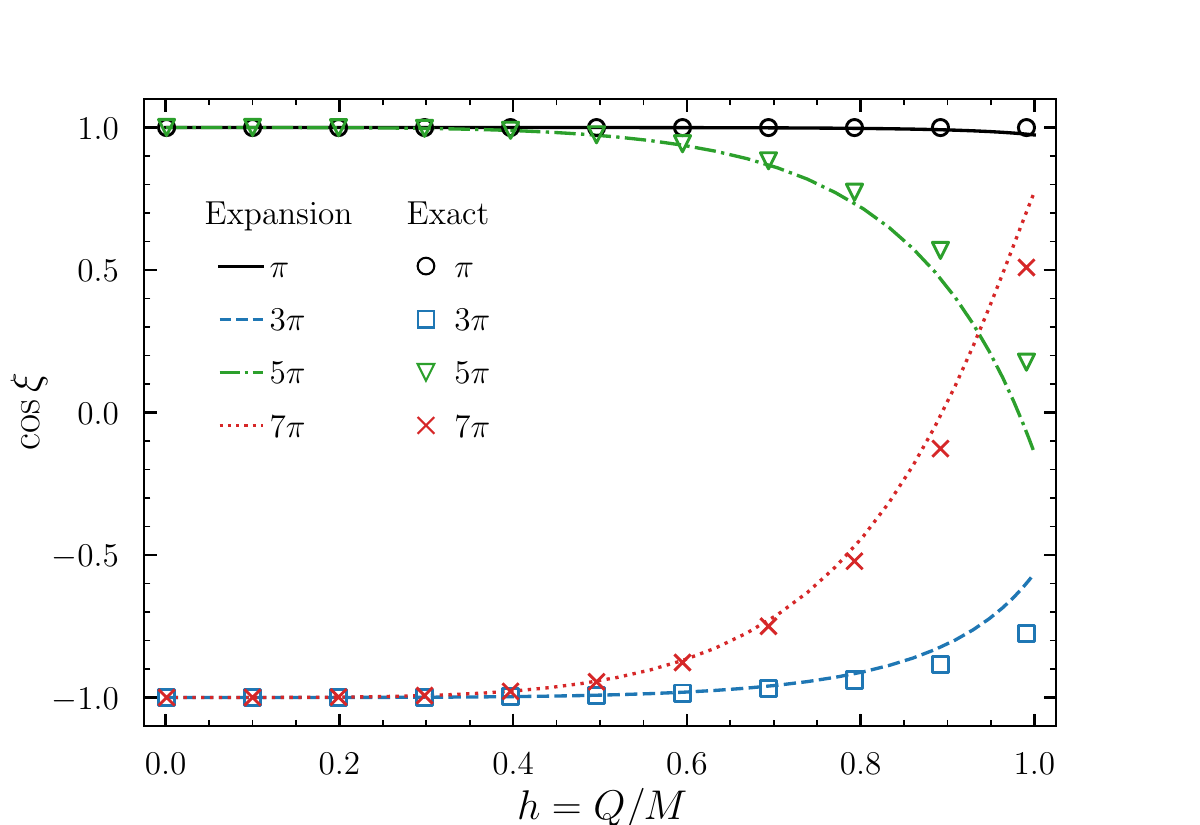}
	\caption{Comparison between the \acrshort{sdl} expansion (curves) and the exact result (markers) of $\cos\xi$ for non-relativistic particles ($v\approx 0$) in the \acrshort{rn} spacetime.
    The line styles for the expansion follow the same convention as in Fig.~\ref{fig:cosxiSch}, while the black circles, blue squares, green triangles, and red crosses mark the exact results for $\Delta\varphi = \pi, 3\pi, 5\pi,$ and $7\pi$, respectively~\cite{Pang:2024tco}.
    The expansion remains accurate even for moderately large values of the charge parameter $h$, validating the perturbative approach.}
	\label{fig:cosxiRNtoexact}
\end{figure}

Fig.~\ref{fig:cosxiRNtoexact} provides a direct comparison between the perturbative \acrshort{sdl} expansion and the exact result of $\cos\xi$~\cite{Pang:2024tco} for non-relativistic particles. The agreement is excellent even for sizable values of $h$, confirming that the $h^2$ perturbative treatment captures the essential charge dependence of spin precession in the strong deflection regime.

\section{Discussion}
\label{sec:discussion}

We have generalized the \acrshort{sdl} analysis to the spin precession angle $\chi$ in general static, spherically symmetric, and asymptotically flat spacetimes. The \acrshort{sdl} coefficients $\lambda_\chi$, $\zeta_{\chi,D}$, and $\zeta_{\chi,R}$ were obtained in closed form, and a simple relation~\eqref{eq:chi0phi0relation} linking $\chi_0$ to the deflection angle $\varphi_0$ was established.

The formalism was applied to Schwarzschild and \acrshort{rn} spacetimes. In the Schwarzschild case, analytic expressions for all coefficients were derived, and as expected the regular part $\zeta_{\chi,R}$ approaches $\pi/2$ in the massless limit, consistent with the universal spin flip $\cos\xi=-1$ and the consequent absence of a glory spot. For the \acrshort{rn} spacetime, a perturbative expansion to order $h^2$ yielded the charge corrections to all \acrshort{sdl} coefficients. While the charge introduces a non-trivial dependence for massive particles that distinguishes the \acrshort{rn} case from Schwarzschild, the spin precession of massless particles remains charge-independent, consistent with the universal spin-flip expectation.

Several extensions are worth pursuing. Firstly, the formalism need to be generalized to metrics with $g_{tt}\neq g_{rr}^{-1}$, which arise naturally in \acrfull{lqg} corrected black holes~\cite{Zhang:2024khj}, and to regular black hole models and modified gravity theories to probe how deviations from general relativity manifest in spin precession. Secondly, extending the analysis to Kerr--NUT--(A)dS spacetimes, where hidden symmetries~\cite{Frolov:2006dqt,Frolov:2008jr} render both the geodesic and parallel transport equations separable~\cite{Connell:2008vn,Kubiznak:2008zs}, would reveal the interplay between spacetime rotation and particle spin in the strong-deflection regime, though the discussion of glory scattering becomes less important in the absence of spherical symmetry. Finally, incorporating the \acrshort{mpd} spin-curvature coupling~\cite{Mathisson:2010opl,Papapetrou:1951pa} would clarify whether the vanishing glory spot for massless spinning particles persists when spin-dependent trajectory modifications are taken into account.

\acknowledgments
The authors appreciate the financial support from the National Natural  Science Foundation of China (NSFC) with Grant Nos. 12303048 and 12505082. XP is also supported by the Doctoral Initiation Fund 24KE051 and the Basic Research Grant 25kx010 from China West Normal University. QJ is supported by the Key Joint Program of Science and Education of Sichuan Province with Grant No. 25LHJJ0097.  YX is supported by the Doctoral Initiation Fund of West China Normal University (22kE040) and the Open Fund of Key Laboratory of Astroparticle Physics of Yunnan Province (2022Zibian3).

\appendix 
\section{The expansion of $r_0$ and $\eta$}
\label{sec:expr0andeta}

In this appendix, we are going to show that for $r_0=r_c(1+\delta)$, we have $\eta=\eta_c(1+ \gamma \delta^2)$ for some constant $\gamma$. To see this, we solve Eqs.~\eqref{eq:rcdefeq} for $v$ and $\eta$ to obtain
\begin{IEEEeqnarray}{rCl}
	v &=& \frac{\sqrt{-2f_c+2f_c^2+r_c f'_c}}{\sqrt{2}f_c} \quad, \quad \eta_c=\frac{r_c^{3/2}\sqrt{f'_c}}{\sqrt{2}f_c} \quad , \label{eq:vfrcrule}
\end{IEEEeqnarray} 
where we have introduced the shorthand notations $f_c=f(r_c)$ and  $f'_c=f'(r_c)$. %
Note that although we have written $v$ as a function of $r_c$, in general we would like to fix $v$ and calculate the corresponding $r_c$ (or $\eta_c$). 
Conversely, we can use $v$ and $\eta_c$ to express $f_c$ and $f'_c$ as
\begin{IEEEeqnarray}{rCl}
	f_c &=& \frac{r_c^2}{\eta_c+r_c^2(1-v^2)} \quad, \quad f'_c=\frac{2\eta_c^2r_c}{\left[\eta_c^2+r_c^2(1-v^2)\right]^2} \quad . \label{eq:gcgpcrule}
\end{IEEEeqnarray} 

For general $\eta$ instead of the critical one $\eta_c$, there is also a corresponding $r_0$, and their relation can be given by solving the equation $\frac{\dd r}{\dd\varphi}=0$ to obtain
\begin{IEEEeqnarray}{rCl}
	\eta &=& \frac{r_0\sqrt{1-(1-v^2)f_0}}{\sqrt{f_0}} \quad , \label{eq:etar0relation}
\end{IEEEeqnarray} 
where we again use the shorthand notation $ f_0 = f (r_0)$.
 Letting $r_0=r_c(1+\delta)$ and expand the right-hand side respect to small $\delta$ to the second order $\delta^2$, then substitute $f_c$ and $f'_c$ given by \eqref{eq:gcgpcrule}, we find the $\delta$ term vanishes and obtain 
\begin{IEEEeqnarray}{rCl}
	\eta &=& \eta_c(1+ \gamma\delta^2)  \quad, 
\end{IEEEeqnarray} 
with 
\begin{IEEEeqnarray}{rCl}
	\gamma &=& \frac{2\eta_c^4-6\eta_c^2r_c^2(1-v^2)-\left[\eta_c^2+r_c^2(1-v^2)\right]^3f''_c}{4\eta_c^2 \left[\eta_c^2+r_c^2(1-v^2)\right]} \quad.  \label{eq:r0etagamma}
\end{IEEEeqnarray} 
In particular, for massless particles with $v=1$, we have 
\begin{IEEEeqnarray}{rCl}
	\gamma|_{v=1} &=& \frac{1}{2}-\frac{\eta_c^2f''_c}{4} \quad.  \label{eq:r0etagammav1}
\end{IEEEeqnarray} 

Conversely, if we let $\eta=\eta_c(1+\varepsilon)$, then we will have 
\begin{IEEEeqnarray}{rCl}
	r_0 &=& r_c\left(1+\sqrt{\frac{\varepsilon}{\gamma}}\right) \quad. \label{eq:r0rcepsilongamma}
\end{IEEEeqnarray} 
This relation is needed if one want to translate the expansion of deflection angle $\varphi$ from $r_c$ to $\eta_c$.

\printglossary[title=Acronyms]


\begin{thebibliography}{10}

\bibitem{Landau:1982dva}
L.D.~Landau and E.M.~Lifschits, \emph{The {{Classical Theory}} of {{Fields}}}, Pergamon Press, Oxford (1975).

\bibitem{Straumann:1984grr}
N.~Straumann, \emph{General {{Relativity}} and {{Relativistic Astrophysics}}}, Springer, Berlin, Heidelberg (1984), \href{https://doi.org/10.1007/978-3-642-84439-3}{10.1007/978-3-642-84439-3}.

\bibitem{Stone:2014fja}
M.~Stone, V.~Dwivedi and T.~Zhou, \emph{Berry phase, {{Lorentz}} covariance, and anomalous velocity for {{ Dirac}} and {{Weyl}} particles}, \href{https://doi.org/10.1103/PhysRevD.91.025004}{\emph{Physical Review D} {\bfseries 91} (2015) 025004}.

\bibitem{Oancea:2022utx}
M.A.~Oancea and A.~Kumar, \emph{Semiclassical analysis of {{Dirac}} fields on curved spacetime}, \href{https://doi.org/10.1103/PhysRevD.107.044029}{\emph{Physical Review D} {\bfseries 107} (2023) 044029} [\href{https://arxiv.org/abs/2212.04414}{{\ttfamily 2212.04414}}].

\bibitem{Ford:1959sd}
K.W.~Ford and J.A.~Wheeler, \emph{Semiclassical description of scattering}, {\emph{Annals of Physics} {\bfseries 7} (1959) 259}.

\bibitem{Zhang:1984vt}
T.~Zhang and C.~{DeWitt-Morette}, \emph{{{WKB Cross Section}} for {{Polarized Glories}} of {{Massless Waves }} in {{Curved Space-Times}}}, \href{https://doi.org/10.1103/PhysRevLett.52.2313}{\emph{Phys. Rev. Lett.} {\bfseries 52} (1984) 2313}.

\bibitem{DeWittMorette:1984pk}
C.~{DeWitt-Morette} and B.L.~Nelson, \emph{Glories - and other degenerate points of the action}, \href{https://doi.org/10.1103/PhysRevD.29.1663}{\emph{Phys. Rev.} {\bfseries D29} (1984) 1663}.

\bibitem{Matzner:1985rjna}
R.A.~Matzner, C.~{DeWitte-Morette}, B.~Nelson and T.~Zhang, \emph{Glory scattering by black holes}, \href{https://doi.org/10.1103/PhysRevD.31.1869}{\emph{Physical Review D} {\bfseries 31} (1985) 1869}.

\bibitem{Futterman:1988sbh}
J.A.H.~Futterman, F.A.~Handler and R.A.~Matzner, \emph{{Scattering from Black Holes}}, Cambridge Monographs on Mathematical Physics, Cambridge University Press (1988), \href{https://doi.org/10.1017/CBO9780511735615}{10.1017/CBO9780511735615}.

\bibitem{Dolan:2006vja}
S.~Dolan, C.~Doran and A.~Lasenby, \emph{Fermion scattering by a {{Schwarzschild}} black hole}, \href{https://doi.org/10.1103/PhysRevD.74.064005}{\emph{Phys. Rev. D} {\bfseries 74} (2006) 064005} [\href{https://arxiv.org/abs/gr-qc/0605031}{{\ttfamily gr-qc/0605031}}].

\bibitem{Crispino:2009xt}
L.C.B.~Crispino, S.R.~Dolan and E.S.~Oliveira, \emph{{Electromagnetic wave scattering by Schwarzschild black holes}}, \href{https://doi.org/10.1103/PhysRevLett.102.231103}{\emph{Phys. Rev. Lett.} {\bfseries 102} (2009) 231103} [\href{https://arxiv.org/abs/0905.3339}{{\ttfamily 0905.3339}}].

\bibitem{Crispino:2009kia}
L.C.B.~Crispino, S.R.~Dolan and E.S.~Oliveira, \emph{Scattering of massless scalar waves by reissner-nordstr\"m black holes}, \href{https://doi.org/10.1103/PhysRevD.79.064022}{\emph{Phys. Rev. D} {\bfseries 79} (2009) 064022} [\href{https://arxiv.org/abs/0904.0999}{{\ttfamily 0904.0999}}].

\bibitem{Macedo:2015qma}
C.F.B.~Macedo, E.S.~de~Oliveira and L.C.B.~Crispino, \emph{{Scattering by regular black holes: Planar massless scalar waves impinging upon a Bardeen black hole}}, \href{https://doi.org/10.1103/PhysRevD.92.024012}{\emph{Phys. Rev. D} {\bfseries 92} (2015) 024012} [\href{https://arxiv.org/abs/1505.07014}{{\ttfamily 1505.07014}}].

\bibitem{Huang:2020bdf}
Y.~Huang and H.~Zhang, \emph{Scattering of massless scalar field by charged dilatonic black holes}, \href{https://doi.org/10.1140/epjc/s10052-020-8228-8}{\emph{Eur. Phys. J. C} {\bfseries 80} (2020) 654} [\href{https://arxiv.org/abs/2006.01388}{{\ttfamily 2006.01388}}].

\bibitem{Crispino:2009xta}
L.C.B.~Crispino, S.R.~Dolan and E.S.~Oliveira, \emph{Electromagnetic wave scattering by {{Schwarzschild}} black holes}, \href{https://doi.org/10.1103/PhysRevLett.102.231103}{\emph{Phys. Rev. Lett.} {\bfseries 102} (2009) 231103} [\href{https://arxiv.org/abs/0905.3339}{{\ttfamily 0905.3339}}].

\bibitem{Pang:2024tco}
X.~Pang, Q.~Jiang, Y.~Xiang and G.~Deng, \emph{The precession of particle spin in spherical symmetric spacetimes}, \href{https://doi.org/10.1140/epjc/s10052-025-13894-8}{\emph{Eur. Phys. J. C} {\bfseries 85} (2025) 193} [\href{https://arxiv.org/abs/2410.04323}{{\ttfamily 2410.04323}}].

\bibitem{Kubiznak:2008zs}
D.~Kubiznak, V.P.~Frolov, P.~Krtous and P.~Connell, \emph{Parallel-propagated frame along null geodesics in higher-dimensional black hole spacetimes}, \href{https://doi.org/10.1103/PhysRevD.79.024018}{\emph{Phys. Rev. D} {\bfseries 79} (2009) 024018} [\href{https://arxiv.org/abs/0811.0012}{{\ttfamily 0811.0012}}].

\bibitem{Connell:2008vn}
P.~Connell, V.P.~Frolov and D.~Kubiznak, \emph{Solving parallel transport equations in the higher-dimensional {{ Kerr-NUT-}}({{A}}){{dS}} spacetimes}, \href{https://doi.org/10.1103/PhysRevD.78.024042}{\emph{Phys. Rev. D} {\bfseries 78} (2008) 024042} [\href{https://arxiv.org/abs/0803.3259}{{\ttfamily 0803.3259}}].

\bibitem{LIGOScientific:2016aoc}
{\scshape LIGO Scientific, Virgo} collaboration, \emph{{Observation of Gravitational Waves from a Binary Black Hole Merger}}, \href{https://doi.org/10.1103/PhysRevLett.116.061102}{\emph{Phys. Rev. Lett.} {\bfseries 116} (2016) 061102} [\href{https://arxiv.org/abs/1602.03837}{{\ttfamily 1602.03837}}].

\bibitem{LIGOScientific:2018mvr}
{\scshape LIGO Scientific, Virgo} collaboration, \emph{{GWTC-1: A Gravitational-Wave Transient Catalog of Compact Binary Mergers Observed by LIGO and Virgo during the First and Second Observing Runs}}, \href{https://doi.org/10.1103/PhysRevX.9.031040}{\emph{Phys. Rev. X} {\bfseries 9} (2019) 031040} [\href{https://arxiv.org/abs/1811.12907}{{\ttfamily 1811.12907}}].

\bibitem{Akiyama:2019bqs}
K.~Akiyama et~al., \emph{First {{M87 Event Horizon Telescope Results}}. {{IV}}. {{Imaging}} the {{Central Supermassive Black Hole}}}, \href{https://doi.org/10.3847/2041-8213/ab0e85}{\emph{Astrophys. J.} {\bfseries 875} (2019) L4}.

\bibitem{EventHorizonTelescope:2019dse}
{\scshape Event Horizon Telescope} collaboration, \emph{{First M87 Event Horizon Telescope Results. I. The Shadow of the Supermassive Black Hole}}, \href{https://doi.org/10.3847/2041-8213/ab0ec7}{\emph{Astrophys. J. Lett.} {\bfseries 875} (2019) L1} [\href{https://arxiv.org/abs/1906.11238}{{\ttfamily 1906.11238}}].

\bibitem{EventHorizonTelescope:2022exc}
{\scshape Event Horizon Telescope} collaboration, \emph{{First Sagittarius A* Event Horizon Telescope Results. IV. Variability, Morphology, and Black Hole Mass}}, \href{https://doi.org/10.3847/2041-8213/ac6736}{\emph{Astrophys. J. Lett.} {\bfseries 930} (2022) L15} [\href{https://arxiv.org/abs/2311.08697}{{\ttfamily 2311.08697}}].

\bibitem{EventHorizonTelescope:2022xqj}
{\scshape Event Horizon Telescope} collaboration, \emph{{First Sagittarius A* Event Horizon Telescope Results. VI. Testing the Black Hole Metric}}, \href{https://doi.org/10.3847/2041-8213/ac6756}{\emph{Astrophys. J. Lett.} {\bfseries 930} (2022) L17} [\href{https://arxiv.org/abs/2311.09484}{{\ttfamily 2311.09484}}].

\bibitem{Darwin:1959gfp}
C.G.~Darwin, \emph{The gravity field of a particle}, \href{https://doi.org/10.1098/rspa.1959.0015}{\emph{Proc. R. Soc. Lond. A} (1959) 180}.

\bibitem{Virbhadra:1999nm}
K.S.~Virbhadra and G.F.R.~Ellis, \emph{Schwarzschild black hole lensing}, \href{https://doi.org/10.1103/PhysRevD.62.084003}{\emph{Phys. Rev. D} {\bfseries 62} (2000) 084003} [\href{https://arxiv.org/abs/astro-ph/9904193}{{\ttfamily astro-ph/9904193}}].

\bibitem{Claudel:2000yi}
C.-M.~Claudel, K.S.~Virbhadra and G.F.R.~Ellis, \emph{The {{Geometry}} of photon surfaces}, \href{https://doi.org/10.1063/1.1308507}{\emph{J. Math. Phys.} {\bfseries 42} (2001) 818} [\href{https://arxiv.org/abs/gr-qc/0005050}{{\ttfamily gr-qc/0005050}}].

\bibitem{Bozza:2002zj}
V.~Bozza, \emph{Gravitational lensing in the strong field limit}, \href{https://doi.org/10.1103/PhysRevD.66.103001}{\emph{Phys. Rev. D} {\bfseries 66} (2002) 103001} [\href{https://arxiv.org/abs/gr-qc/0208075}{{\ttfamily gr-qc/0208075}}].

\bibitem{Iyer:2006cn}
S.V.~Iyer and A.O.~Petters, \emph{Light's bending angle due to black holes: From the photon sphere to infinity}, \href{https://doi.org/10.1007/s10714-007-0481-8}{\emph{General Relativity and Gravitation} {\bfseries 39} (2007) 1563} [\href{https://arxiv.org/abs/gr-qc/0611086}{{\ttfamily gr-qc/0611086}}].

\bibitem{Tsupko:2014wza}
O.r.Y.~Tsupko, \emph{Unbound motion of massive particles in the {{Schwarzschild}} metric: {{Analytical}} description in case of strong deflection}, \href{https://doi.org/10.1103/PhysRevD.89.084075}{\emph{Phys. Rev. D} {\bfseries 89} (2014) 084075} [\href{https://arxiv.org/abs/1505.06481}{{\ttfamily 1505.06481}}].

\bibitem{Pang:2018jpm}
X.~Pang and J.~Jia, \emph{Gravitational lensing of massive particles in {{Reissner}}–{{ Nordström}} black hole spacetime}, \href{https://doi.org/10.1088/1361-6382/ab0512}{\emph{Class. Quant. Grav.} {\bfseries 36} (2019) 065012} [\href{https://arxiv.org/abs/1806.04719}{{\ttfamily 1806.04719}}].

\bibitem{Feleppa:2024kio}
F.~Feleppa, V.~Bozza and O.Y.~Tsupko, \emph{Strong deflection of massive particles in spherically symmetric spacetimes}, \href{https://doi.org/10.1103/PhysRevD.111.044018}{\emph{Phys. Rev. D} {\bfseries 111} (2025) 044018} [\href{https://arxiv.org/abs/2412.16712}{{\ttfamily 2412.16712}}].

\bibitem{Weinberg:1972kfs}
S.~Weinberg, \emph{Gravitation and {{Cosmology}}}, {John Wiley and Sons}, New York (1972).

\bibitem{Khriplovich:2008ni}
I.B.~Khriplovich, \emph{Spinning relativistic particles in external fields},  Jan., 2008.

\bibitem{Dolan:2017zgu}
S.R.~Dolan, \emph{Geometrical optics for scalar, electromagnetic and gravitational waves in curved spacetime}, \href{https://doi.org/10.1142/S0218271818430101}{\emph{Int. J. Mod. Phys. D} {\bfseries 27} (2018) 1843010} [\href{https://arxiv.org/abs/1806.08617}{{\ttfamily 1806.08617}}].

\bibitem{Mathisson:2010opl}
M.~Mathisson, \emph{Republication of: {{New}} mechanics of material systems}, \href{https://doi.org/10.1007/s10714-010-0939-y}{\emph{Gen. Rel. Grav.} {\bfseries 42} (2010) 1011}.

\bibitem{Papapetrou:1951pa}
A.~Papapetrou, \emph{Spinning test particles in general relativity. 1.}, \href{https://doi.org/10.1098/rspa.1951.0200}{\emph{Proc. Roy. Soc. Lond. A} {\bfseries 209} (1951) 248}.

\bibitem{Rudiger:1981uu}
R.~Rudiger, \emph{The {{Dirac}} equation and spinning particles in general relativity}, \href{https://doi.org/10.1098/rspa.1981.0132}{\emph{Proc. Roy. Soc. Lond. A} {\bfseries 377} (1981) 417}.

\bibitem{Mohseni:2010rm}
M.~Mohseni, \emph{Stability of circular orbits of spinning particles in {{ Schwarzschild-like}} space-times}, \href{https://doi.org/10.1007/s10714-010-0995-3}{\emph{Gen. Rel. Grav.} {\bfseries 42} (2010) 2477} [\href{https://arxiv.org/abs/1005.3110}{{\ttfamily 1005.3110}}].

\bibitem{Zhang:2017nhl}
Y.-P.~Zhang, S.-W.~Wei, W.-D.~Guo, T.-T.~Sui and Y.-X.~Liu, \emph{Innermost stable circular orbit of spinning particle in charged spinning black hole background}, \href{https://doi.org/10.1103/PhysRevD.97.084056}{\emph{Phys. Rev. D} {\bfseries 97} (2018) 084056} [\href{https://arxiv.org/abs/1711.09361}{{\ttfamily 1711.09361}}].

\bibitem{Iorio:2011ubn}
L.~Iorio, \emph{General relativistic spin-orbit and spin-spin effects on the motion of rotating particles in an external gravitational field}, \href{https://doi.org/10.1007/s10714-011-1302-7}{\emph{Gen. Rel. Grav.} {\bfseries 44} (2012) 719} [\href{https://arxiv.org/abs/1012.5622}{{\ttfamily 1012.5622}}].

\bibitem{Chakraborty:2016mhx}
C.~Chakraborty, M.~Patil, P.~Kocherlakota, S.~Bhattacharyya, P.S.~Joshi and A.~Kr{\'o}~lak, \emph{Distinguishing {{Kerr}} naked singularities and black holes using the spin precession of a test gyro in strong gravitational fields}, \href{https://doi.org/10.1103/PhysRevD.95.084024}{\emph{Phys. Rev. D} {\bfseries 95} (2017) 084024} [\href{https://arxiv.org/abs/1611.08808}{{\ttfamily 1611.08808}}].

\bibitem{Witzany:2023bmq}
V.~Witzany and G.A.~Piovano, \emph{Analytic {{Solutions}} for the {{Motion}} of {{Spinning Particles}} near {{Spherically Symmetric Black Holes}} and {{Exotic Compact Objects}}}, \href{https://doi.org/10.1103/PhysRevLett.132.171401}{\emph{Phys. Rev. Lett.} {\bfseries 132} (2024) 171401} [\href{https://arxiv.org/abs/2308.00021}{{\ttfamily 2308.00021}}].

\bibitem{Witzany:2026eqc}
V.~Witzany and V.~Skoup{\'y}, \emph{{Separability of the motion of spinning test particles in curved space-time}},  \href{https://arxiv.org/abs/2606.25792}{{\ttfamily 2606.25792}}.

\bibitem{Frolov:2017kze}
V.P.~Frolov, P.~Krtous and D.~Kubiznak, \emph{{Black holes, hidden symmetries, and complete integrability}}, \href{https://doi.org/10.1007/s41114-017-0009-9}{\emph{Living Rev. Rel.} {\bfseries 20} (2017) 6} [\href{https://arxiv.org/abs/1705.05482}{{\ttfamily 1705.05482}}].

\bibitem{Poisson:2011nh}
E.~Poisson, A.~Pound and I.~Vega, \emph{The motion of point particles in curved spacetime}, \href{https://doi.org/10.12942/lrr-2011-7}{\emph{Living Rev. Rel.} {\bfseries 14} (2011) 7} [\href{https://arxiv.org/abs/1102.0529}{{\ttfamily 1102.0529}}].

\bibitem{Duval:2016hxo}
C.~Duval and T.~Schucker, \emph{Gravitational birefringence of light in {{Robertson-Walker}} cosmologies}, \href{https://doi.org/10.1103/PhysRevD.96.043517}{\emph{Phys. Rev. D} {\bfseries 96} (2017) 043517} [\href{https://arxiv.org/abs/1610.00555}{{\ttfamily 1610.00555}}].

\bibitem{Frolov:2020uhn}
V.P.~Frolov, \emph{Maxwell equations in a curved spacetime: {{Spin}} optics approximation}, \href{https://doi.org/10.1103/PhysRevD.102.084013}{\emph{Phys. Rev. D} {\bfseries 102} (2020) 084013} [\href{https://arxiv.org/abs/2007.03743}{{\ttfamily 2007.03743}}].

\bibitem{Shoom:2020zhr}
A.A.~Shoom, \emph{Gravitational {{Faraday}} and spin-{{Hall}} effects of light}, \href{https://doi.org/10.1103/PhysRevD.104.084007}{\emph{Phys. Rev. D} {\bfseries 104} (2021) 084007} [\href{https://arxiv.org/abs/2006.10077}{{\ttfamily 2006.10077}}].

\bibitem{Frolov:2024ebe}
V.P.~Frolov, \emph{Spinoptics in a curved spacetime}, \href{https://doi.org/10.1103/PhysRevD.110.064020}{\emph{Phys. Rev. D} {\bfseries 110} (2024) 064020} [\href{https://arxiv.org/abs/2405.01777}{{\ttfamily 2405.01777}}].

\bibitem{Takeuchi:2026pyi}
T.~Takeuchi and T.~Kobayashi, \emph{{Spinoptics in the presence of axion-like particles in curved spacetime}},  \href{https://arxiv.org/abs/2607.11134}{{\ttfamily 2607.11134}}.

\bibitem{Andersson:2020gsj}
L.~Andersson, J.~Joudioux, M.A.~Oancea and A.~Raj, \emph{Propagation of polarized gravitational waves}, \href{https://doi.org/10.1103/PhysRevD.103.044053}{\emph{Phys. Rev. D} {\bfseries 103} (2021) 044053} [\href{https://arxiv.org/abs/2012.08363}{{\ttfamily 2012.08363}}].

\bibitem{Frolov:2024qow}
V.P.~Frolov and A.A.~Shoom, \emph{Gravitational spinoptics in a curved space-time}, \href{https://doi.org/10.1088/1475-7516/2024/10/039}{\emph{JCAP} {\bfseries 10} (2024) 039} [\href{https://arxiv.org/abs/2406.17905}{{\ttfamily 2406.17905}}].

\bibitem{Harte:2018wni}
A.I.~Harte, \emph{Gravitational lensing beyond geometric optics: {{I}}. {{Formalism}} and observables}, \href{https://doi.org/10.1007/s10714-018-2494-x}{\emph{Gen. Rel. Grav.} {\bfseries 51} (2019) 14} [\href{https://arxiv.org/abs/1808.06203}{{\ttfamily 1808.06203}}].

\bibitem{Oancea:2020khc}
M.A.~Oancea, J.~Joudioux, I.Y.~Dodin, D.E.~Ruiz, C.F.~Paganini and L.~Andersson, \emph{Gravitational spin {{Hall}} effect of light}, \href{https://doi.org/10.1103/PhysRevD.102.024075}{\emph{Phys. Rev. D} {\bfseries 102} (2020) 024075} [\href{https://arxiv.org/abs/2003.04553}{{\ttfamily 2003.04553}}].

\bibitem{Oancea:2023ylb}
M.A.~Oancea and T.~Harko, \emph{Weyl geometric effects on the propagation of light in gravitational fields}, \href{https://doi.org/10.1103/PhysRevD.109.064020}{\emph{Phys. Rev. D} {\bfseries 109} (2024) 064020} [\href{https://arxiv.org/abs/2305.01313}{{\ttfamily 2305.01313}}].

\bibitem{Landau:mechanics}
L.D.~Landau and E.M.~Lifshits, \emph{Mechanics}, Pergamon Press, Oxford (1969).

\bibitem{Liu:2015zou}
X.~Liu, N.~Yang and J.~Jia, \emph{Gravitational lensing of massive particles in {{Schwarzschild}} gravity}, \href{https://doi.org/10.1088/0264-9381/33/17/175014}{\emph{Class. Quant. Grav.} {\bfseries 33} (2016) 175014} [\href{https://arxiv.org/abs/1512.04037}{{\ttfamily 1512.04037}}].

\bibitem{Zhang:2024khj}
C.~Zhang, J.~Lewandowski, Y.~Ma and J.~Yang, \emph{{Black holes and covariance in effective quantum gravity}}, \href{https://doi.org/10.1103/PhysRevD.111.L081504}{\emph{Phys. Rev. D} {\bfseries 111} (2025) L081504} [\href{https://arxiv.org/abs/2407.10168}{{\ttfamily 2407.10168}}].

\bibitem{Frolov:2006dqt}
V.P.~Frolov and D.~Kubiznak, \emph{`{{Hidden}}' {{Symmetries}} of {{Higher Dimensional Rotating Black Holes}}}, \href{https://doi.org/10.1103/PhysRevLett.98.011101}{\emph{Physical Review Letters} {\bfseries 98} (2007) 011101} [\href{https://arxiv.org/abs/gr-qc/0605058}{{\ttfamily gr-qc/0605058}}].

\bibitem{Frolov:2008jr}
V.P.~Frolov and D.~Kubiznak, \emph{Higher-{{Dimensional Black Holes}}: {{Hidden Symmetries}} and {{ Separation}} of {{Variables}}}, \href{https://doi.org/10.1088/0264-9381/25/15/154005}{\emph{Classical and Quantum Gravity} {\bfseries 25} (2008) 154005} [\href{https://arxiv.org/abs/0802.0322}{{\ttfamily 0802.0322}}].

\end{thebibliography}

\providecommand{\href}[2]{#2}\begingroup\raggedright\endgroup

\end{document}